\documentclass[12pt,english]{article}
\usepackage[T1]{fontenc}
\usepackage[utf8]{inputenc}
\usepackage{babel}
\usepackage{array}
\usepackage{longtable}
\usepackage{amsmath}
\usepackage{amsthm}
\usepackage{amssymb}
\usepackage{geometry}
\usepackage{setspace}
\usepackage[authoryear]{natbib}
\usepackage[pdfusetitle,
 bookmarks=true,bookmarksnumbered=true,bookmarksopen=true,bookmarksopenlevel=2,
 breaklinks=false,pdfborder={0 0 1},backref=false,colorlinks=false]
 {hyperref}

\makeatletter

\providecommand{\tabularnewline}{\\}

\theoremstyle{plain}
\newtheorem{prop}{\protect\propositionname}
\theoremstyle{plain}
\newtheorem{thm}{\protect\theoremname}
\theoremstyle{plain}
\newtheorem{cor}{\protect\corollaryname}
\theoremstyle{remark}
\newtheorem{rem}{\protect\remarkname}
\theoremstyle{definition}
 \newtheorem{example}{\protect\examplename}

\makeatletter
\AtBeginDocument{%
\@ifundefined{c@lem}{}{}%
\@ifundefined{c@lemma}{}{}}
\makeatother
\usepackage{tikz}
\usetikzlibrary{arrows.meta,positioning,calc}
\newcommand{\roadmapfig}{%
\begin{tikzpicture}[
  font=\footnotesize, >=Stealth, x=1cm, y=1cm,
  box/.style={draw, rounded corners=2pt, align=center, inner sep=3pt, minimum height=10mm, text width=30mm},
  asm/.style={box, fill=gray!15},
  res/.style={box, line width=0.7pt},
  imp/.style={->, thick},
  sub/.style={->, thick, dashed},
  cmp/.style={<->, thick, dotted},
  lab/.style={font=\scriptsize, fill=white, inner sep=1.5pt}]
\def\xa{0}\def\xb{3.85}\def\xc{7.7}\def\xd{11.55}
\node[asm] (Db) at (\xa,0)  {D(b)\\ Markov errors};
\node[asm] (Dc) at (\xb,0)  {D(c)\\ $\beta$-mixing errors};
\node[asm] (Dd) at (\xc,0)  {D(d)\\ $\alpha$-mixing errors};
\draw[sub] (Db) -- (Dc);
\draw[sub] (Dc) -- (Dd);
\node[res] (P2)  at (\xa,-2.2) {Prop.~3 (+ D(a))\\ A3 with $C_D=O(1)$};
\node[res] (P3a) at (\xb,-2.2) {Prop.~4(a) (+ D(a))\\ blocking, coupling};
\node[res] (P3b) at (\xc,-2.2) {Prop.~4(b) (+ D(a))\\ Bernstein, grouping};
\draw[imp] (Db) -- (P2);
\draw[imp] (Dc) -- (P3a);
\draw[imp] (Dd) -- (P3b);
\node[asm] (A3) at (\xa,-4.7)  {A3\\ concentration, $C_D$};
\node[res] (P1) at (\xb,-4.7)  {Prop.~1\\ sub-Gaussian increments};
\node[res] (T1) at (\xc,-4.7)  {\textbf{Theorem 1}\\ consistency, $L_{NT}/\sqrt{C_D}$};
\node[res] (T3) at (\xd,-4.7)  {Theorem 3\\ minimax lower bound:\\ rate is optimal};
\draw[imp] (P2) -- (A3);
\draw[imp] (A3) -- (P1);
\draw[imp] (P1) -- (T1);
\draw[imp] (P3a.south east) -- node[lab, pos=0.55, left=3pt] {$L_{NT}/\sqrt{\log NT}$} (T1.north west);
\draw[imp] (P3b) -- node[lab, right] {$L_{NT}/(\log NT\,\log T)$} (T1);
\draw[sub] (T3) -- (T1);
\node[res] (P6) at (\xa,-7.0) {Prop.~2\\ A3 vs.\ eigenvalue\\ condition};
\draw[cmp] (A3) -- node[lab, right] {not nested} (P6);
\node[res] (C1) at (\xb,-7.0) {Corollary 1\\ number of factors};
\node[res] (T2) at (\xc,-7.0) {\textbf{Theorem 2}\\ asymptotic normality};
\node[asm] (A4) at (\xd,-7.0) {A4, Assumption~S\\ mixing, smoothing};
\draw[imp] (T1) -- (C1);
\draw[imp] (T1) -- node[lab, right] {$C_D=O(1)$} (T2);
\draw[imp] (A4) -- (T2);
\node[res] (P4) at (5.775,-9.3) {Prop.~5\\ HAC variance estimation};
\node[res] (P5) at (9.625,-9.3) {Prop.~6\\ thresholded $\widehat\Omega^{f}$};
\draw[imp] (T2) -- (P4);
\draw[imp] (T2) -- (P5);
\draw[imp] (P4) -- (P5);
\node[res] (C2) at (5.775,-11.5) {Corollary 2\\ data-driven $B_T$, $b$};
\draw[imp] (P4) -- (C2);
\node[res] (P7) at (\xa,-9.3) {Prop.~7\\ conditional inference};
\draw[imp] (P6) -- (P7);
\draw[imp] (P4) -- (P7);
\node[font=\scriptsize, align=left, anchor=west] at (8.3,-11.5) {%
\tikz[baseline=-0.6ex]{\draw[imp](0,0)--(0.6,0);}\ used in\\[3pt]
\tikz[baseline=-0.6ex]{\draw[sub](0,0)--(0.6,0);}\ implies or bounds\\[3pt]
\tikz[baseline=-0.6ex]{\draw[cmp](0,0)--(0.6,0);}\ compared with\\[3pt]
\tikz[baseline=-0.6ex]{\node[draw, rounded corners=1pt, fill=gray!15, minimum width=6mm, minimum height=3mm]{};}\ assumption};
\end{tikzpicture}%
}

\usepackage{etoolbox}
\AtBeginEnvironment{table}{\setstretch{1}}

\AtBeginEnvironment{table}{\setlength{\tabcolsep}{5pt}}

\makeatother

\providecommand{\corollaryname}{Corollary}
\providecommand{\examplename}{Example}
\providecommand{\propositionname}{Proposition}
\providecommand{\remarkname}{Remark}
\providecommand{\theoremname}{Theorem}

\begin{document}
\title{Approximate Quantile Factor Models}
\author{Xiaochun Liu\thanks{Department of Economics, Finance and Legal Studies, The Culverhouse
College of Business, The University of Alabama, Tuscaloosa, Alabama,
35487, USA. Email: xliu121@ua.edu}}
\maketitle
\begin{abstract}
Weakly dependent idiosyncratic errors are standard in mean factor
models, but quantile factor models have been analyzed only under independence.
The mean-case definition of weak dependence, a bound on the eigenvalues
of the error covariance matrix, does not carry over: I show that it
neither implies nor is implied by the property that controls quantile
estimation, since uncorrelated errors can share a pervasive quantile
component. I instead define approximate quantile factor models through
a concentration condition on the idiosyncratic array. Under it, the
quantile factor analysis (QFA) estimator is consistent at a rate that
deteriorates by the square root of a dependence constant, which is
minimax optimal, and the number of factors is consistently estimated.
Dependence changes the asymptotic variance, for which I provide HAC
and thresholded estimators. Monte Carlo experiments confirm these
predictions. In FRED-QD and FRED-MD, standard errors ignoring dependence
understate uncertainty, most strongly in the tails.


\end{abstract}
\newpage{}

\section{Introduction}

Large-dimensional factor models summarize the comovement of many series
through a few common factors and are a workhorse of macroeconomic
and financial modeling (Stock and Watson, 2002; Bai and Ng, 2002).
A foundational distinction in that literature is between \emph{exact}
and \emph{approximate} factor models (Chamberlain and Rothschild,
1983): the exact model requires the idiosyncratic components to be
mutually uncorrelated, whereas the approximate model permits weak
cross-sectional (and serial) dependence among them, with the common
part distinguished by diverging eigenvalues. The move from exact to
approximate is what made factor models usable on real data, where
idiosyncratic terms are never perfectly uncorrelated; Bai and Ng (2002)
establish consistent estimation and factor-number selection for approximate
\emph{static} factor models of the conditional mean, and this approximate-error
robustness is now the default setting of the applied literature.

Mean factor models are, however, silent about how comovement varies
across the conditional distribution: whether the factors that drive
the center also drive the tails, and whether the number of pervasive
factors differs across quantiles. These features matter for distributional
and tail-risk analysis. Chen, Dolado, and Gonzalo (2021, henceforth
CDG) introduce the quantile factor model and its estimator, quantile
factor analysis (QFA), which minimizes the check-function loss and
allows a quantile-dependent number of factors $r(\tau)$. Korobilis
and Schröder (2025b) develop a probabilistic counterpart estimated
by variational Bayes, and Korobilis and Schröder (2025a) embed quantile
factors in a factor-augmented vector autoregression (QFAVAR) to monitor
macroeconomic risk. All of these work with exact factor models, in
which the idiosyncratic errors are independent across units and time
(CDG, Assumption 1(iii)), an assumption that CDG themselves describe
as strong (their Remark 1.4). The quantile factor literature thus
has no counterpart of the approximate factor model.

What, then, is an approximate quantile factor model? In the mean case,
the answer is a bound on the largest eigenvalue of the idiosyncratic
covariance matrix $\mathbb{E}[u_{t}u_{t}']$, a second-moment object.
At a quantile, the idiosyncratic components enter the check-function
criterion only through indicators $\mathbb{I}\{u_{it}\le s\}$. At
the true parameters the threshold is $s=0$, which gives the score
$\psi_{it}(\tau)=\tau-\mathbb{I}\{u_{it}(\tau)\le0\}$, but at any
other candidate value the thresholds $s$ range over the gap between
the candidate and the true common component. The criterion must therefore
be controlled through such bounded functions of the errors, and uniformly
over a parameter space of dimension $(N+T)r$. The covariance of the
errors is not the relevant object: Proposition 2 shows that the eigenvalue
condition neither implies nor is implied by the concentration property
on which the theory rests. Uncorrelated errors can share a pervasive
quantile component that second moments cannot detect: if a common
shock moves the $\tau$-quantile of every error while leaving all
means and covariances unchanged, the eigenvalue condition holds, yet
QFA can be inconsistent, because the shock acts as an omitted quantile
factor. The answer must therefore be sought in the check-function
criterion itself.

The question matters in practice. In the FRED-QD and FRED-MD panels,
which are standard in the factor literature, the QFA scores are serially
correlated for 40 to 90 percent of the series, and far more pairs
of series are cross-sectionally correlated than independence allows
(Section 5). Under such dependence the confidence intervals of CDG
are invalid. In the simulations of Section 4, the coverage of their
nominal 95 percent intervals for the loadings falls to 0.59 when the
errors follow an autoregression with coefficient 0.8. In the FRED
panels, the standard errors of the factors that account for dependence
are 1.4 to 1.6 times those that assume independence in the tails of
the distribution, where quantile factor models are most often used.

\emph{Contribution.} This paper defines and analyzes the approximate
quantile factor model, and its contributions are four. First, it formulates
weak dependence at a quantile as a concentration property of the idiosyncratic
array, with a dependence constant $C_{D}$ (Assumption A3); the exact
model is the case $C_{D}=1/2$, and primitive conditions cover Markov-driven
and mixing errors (Propositions 3 and 4). Second, it shows that consistency
reduces to concentration: QFA, without modification, estimates the
factors and the loadings at the rate $L_{NT}/\sqrt{C_{D}}$, with
$L_{NT}=\min\{\sqrt{N},\sqrt{T}\}$, and the number of factors consistently
(Theorem 1 and Corollary 1). Because the concentration bound may hold
conditional on a common volatility path, the theory also covers dependence
that A3 excludes (Proposition 2).

Third, it derives the limiting distribution under dependence, shows
that the variance estimators of CDG are then inconsistent, and provides
consistent HAC and thresholded estimators (Theorem 2, Propositions
5 and 6, and Corollary 2), which remain valid conditional on common
volatility (Proposition 7). Fourth, it shows that the rate is sharp:
a minimax lower bound of order $\sqrt{C_{D}}/L_{NT}$ implies that
dependence costs exactly the factor $\sqrt{C_{D}}$ and that uniform
consistency requires $C_{D}=o(L_{NT}^{2})$ (Theorem 3). Together,
these results play the role for quantile factor models that Bai and
Ng (2002) and Bai (2003) play for mean factor models, but the notion
of weak dependence on which they rest has no counterpart in the mean
case.

These results require new arguments rather than a modification of
the exact-model proofs. CDG note that their consistency proof uses
independence only to derive a sub-Gaussian bound for the criterion;
the difficulty lies in finding dependence conditions under which such
a bound holds. The bound must hold uniformly over the parameter space,
for a nonsmooth loss and with the heterogeneous weights that arise
in the criterion, and several standard concentration inequalities
for dependent data fail this requirement (Section 3.1). The distribution
theory requires control of quadratic terms in the expansion that vanish
under independence, as well as long-run variance estimators in which
the estimated scores enter through indicator functions, and the lower
bound requires a construction in which the dependence constant is
explicit.

This paper treats the static quantile factor model, the setting of
Bai and Ng (2002) in the mean case, so that the role of idiosyncratic
dependence is isolated. The results are stated for QFA, but because
the asymmetric Laplace working likelihood of Korobilis and Schröder
(2025b) corresponds to the check-function criterion, the concentration
argument offers a natural route to weakly dependent errors in their
setting as well. Extending the approximate theory to dynamic quantile
factor models, in which the factors follow a vector autoregression
as in the QFAVAR of Korobilis and Schröder (2025a), is left for future
research.

The paper proceeds as follows. Section 2 sets out the quantile factor
model, the exact and approximate assumptions, and the QFA estimator.
Section 3 states the concentration condition that defines the approximate
quantile factor model, proves consistency under it, compares the concentration
condition with the eigenvalue condition of the mean case, gives primitive
sufficient conditions, derives the limiting distribution with consistent
variance estimators, and shows that the rates are minimax optimal.
Section 4 reports Monte Carlo evidence on the predictions of Section
3, Section 5 applies the results to the FRED-QD and FRED-MD macroeconomic
panels, and Section 6 concludes.

\section{The Quantile Factor Model}

This section presents the model, the exact and approximate quantile
factor models, which differ only in their assumption on the idiosyncratic
components, and the estimator. The identification conditions, which
are common to both, follow, and the section closes with the QFA estimator.

\subsection{Model}

Let $x_{it}$ denote the observation on variable $i=1,\dots,N$ in
period $t=1,\dots,T$, and fix a quantile level $\tau\in(0,1)$. The
quantile factor model specifies the conditional $\tau$-quantile of
$x_{it}$ as a linear combination of $r(\tau)$ latent common factors:
\begin{equation}
Q_{\tau}\!\left(x_{it}\mid\lambda_{i}(\tau),f_{t}(\tau)\right)=\lambda_{i}(\tau)'f_{t}(\tau),\qquad x_{it}=\lambda_{i}(\tau)'f_{t}(\tau)+u_{it}(\tau),\label{eq:qfm}
\end{equation}
with $P\{u_{it}(\tau)\le0\mid F_{0}\}=\tau$, where $F_{0}=(f_{01},\dots,f_{0T})'$
denotes the true factor path. The quantile restriction is imposed
conditional on the entire path, which amounts to strict exogeneity
of the factors with respect to the idiosyncratic quantile restriction.
Identification requires this form once $u_{it}$ is serially dependent.
Equivalently, $F_{0}$ may be treated as nonrandom, as in CDG, and
all probability statements below are conditional on $F_{0}$. In matrix
form, $x_{t}=\Lambda f_{t}+u_{t}$ with $\Lambda=(\lambda_{1},\dots,\lambda_{N})'$.
The factors, the loadings, and the number of factors may all depend
on $\tau$. Every object is therefore quantile-specific, estimation
is carried out one quantile at a time, and I suppress $(\tau)$ when
no confusion arises. The model is \emph{static} in the sense that
no law of motion is imposed on $f_{t}$. Finally, the quantile restriction
in $(\ref{eq:qfm})$ constrains each $u_{it}$ separately and is silent
on how the idiosyncratic components depend on one another.

\subsection{Exact and approximate idiosyncratic structure}

The exact and approximate models differ only in what they assume about
this dependence, that is, about the joint distribution of $u_{t}=(u_{1t},\dots,u_{Nt})'$.
Both assumptions are conveniently expressed in terms of the check-function
score
\[
\psi_{it}(\tau)=\tau-\mathbb{I}\{u_{it}(\tau)\le0\},\qquad\mathbb{E}\!\left[\psi_{it}(\tau)\mid F_{0}\right]=0,
\]
which plays the role of the regression residual in quantile estimation
and is the channel through which the idiosyncratic components enter
the estimator.
\begin{description}
\item [{(Exact)}] The idiosyncratic components are mutually independent
across $i$ and $t$ (CDG, Assumption 1(iii)). In particular, the
score field $\{\psi_{it}(\tau)\}$ is independent across $i$ and
$t$. This is the quantile counterpart of the exact factor model of
Chamberlain and Rothschild (1983), and it is the setting of the existing
quantile factor literature.
\item [{(Approximate)}] The idiosyncratic components are weakly dependent.
In the mean factor model, weak dependence requires the largest eigenvalue
of $\mathbb{E}[u_{t}u_{t}']$ to remain bounded as $N\to\infty$,
while the eigenvalues of the common component diverge (Chamberlain
and Rothschild, 1983; Bai and Ng, 2002). At a quantile the covariance
matrix of the errors is not the relevant object. Instead, I state
the approximate condition as a concentration property of the idiosyncratic
array (Assumption A3 in Section 3.1): sums of bounded functions of
the $u_{it}$ concentrate as they would under independence, up to
a dependence constant $C_{D}$. Weak dependence means $C_{D}=O(1)$,
and the exact model is the special case $C_{D}=1/2$.
\end{description}
Nor does a second-moment condition on the score at the true quantile,
such as summability of $|\mathrm{Cov}(\psi_{it},\psi_{js})|$, suffice
for consistency, for two reasons. First, the criterion must be controlled
uniformly over a parameter set of dimension $(N+T)r$, which requires
exponential tail bounds. Second, at parameter values other than the
true one, the criterion involves the indicators $\mathbb{I}\{u_{it}\le s\}$
at nonzero thresholds $s$, given by the gap between the candidate
and the true common component, and a condition on the score $\psi_{it}$
alone says nothing about them. Summability of the score covariances
remains the relevant condition for the asymptotic variance (Section
3.4). Section 3.2 justifies this definition: Proposition 2 there shows
that A3 and the eigenvalue condition are not nested, and that A3 excludes
pervasive quantile components in the idiosyncratic term that the eigenvalue
condition permits. It also shows that A3 may be imposed conditional
on a larger $\sigma$-field that preserves the quantile restriction,
such as one generated by a common volatility path.

The two assumptions differ in their treatment of the dependence that
remains after conditioning on the factors. Both are imposed conditional
on the factor path, so dependence induced \emph{through the common
factors} is allowed in either case. The exact assumption rules out
any further dependence, whereas the approximate assumption permits
it. Under either assumption, the common component must be identified
and must dominate the idiosyncratic component. I turn to these conditions
next.

\subsection{Identification and pervasiveness}

The common component $\lambda_{i}'f_{t}$ is invariant to nonsingular
rotations, so the factors and loadings are identified only up to rotation,
and the object of interest is the factor \emph{space}. I use the standard
normalization (Bai and Ng, 2002; CDG): $T^{-1}\sum_{t}f_{t}f_{t}'=I_{r}$,
and $N^{-1}\sum_{i}\lambda_{i}\lambda_{i}'$ is diagonal with nonincreasing
diagonal elements. To separate the factor space from the idiosyncratic
components, I also require the factors to be pervasive (strong).
\begin{description}
\item [{(P)~Pervasiveness.}] $N^{-1}\sum_{i}\lambda_{0i}\lambda_{0i}'\to\mathrm{diag}(\sigma_{1},\dots,\sigma_{r})$
with $\infty>\sigma_{1}>\dots>\sigma_{r}>0$ (CDG, Assumption 1(i)),
and the loadings lie in a compact set.
\end{description}
Condition (P) separates the common component, whose contribution to
cross-sectional averages is $O(1)$, from the idiosyncratic component,
whose contribution vanishes. It also has a familiar leverage interpretation.
Let $M_{\Lambda}=I_{N}-\Lambda(\Lambda'\Lambda)^{-1}\Lambda'$ be
the residual-maker matrix. Its $i$th diagonal element is $1-b_{ii}/N$,
where $b_{ii}/N=\lambda_{i}'(\Lambda'\Lambda)^{-1}\lambda_{i}=O(N^{-1})\to0$
under (P), so no single series has nonvanishing leverage. Condition
(P) is maintained under both the exact and the approximate model.
The number of factors $r(\tau)$ is selected by the rank-minimization
or information criteria of CDG, which are consistent under the approximate
model (Corollary 1).

\subsection{The QFA estimator}

I estimate the model by the quantile factor analysis (QFA) estimator
of CDG. Given $r=r(\tau)$, the factors and loadings jointly minimize
the check-function loss
\begin{equation}
(\widehat{\Lambda},\widehat{F})=\arg\min_{\Lambda,F}\;\frac{1}{NT}\sum_{i=1}^{N}\sum_{t=1}^{T}\rho_{\tau}\!\left(x_{it}-\lambda_{i}'f_{t}\right),\qquad\rho_{\tau}(u)=u\left(\tau-\mathbb{I}\{u\le0\}\right),\label{eq:qfa}
\end{equation}
subject to the normalization of Section 2.3. The objective is not
jointly convex in $(\Lambda,F)$, but it is convex in $\Lambda$ given
$F$ and in $F$ given $\Lambda$. The iterative quantile regression
(IQR) algorithm of CDG exploits this structure by alternating between
the two convex problems until it reaches a stationary point. The criterion
adds up the $(i,t)$ cells as separate quantile regression contributions
and does not model the idiosyncratic dependence, just as principal
components do not in the approximate mean factor model. The question
is which dependence the criterion tolerates. Answering it requires
a condition that controls the criterion uniformly over the parameter
set without independence, and Section 3 provides one.

\section{Asymptotic Theory under Weak Dependence}

This section establishes the asymptotic properties of QFA under the
approximate model. Section 3.1 states the assumptions and establishes
consistency of the estimated factors, loadings, and number of factors.
Section 3.2 shows that A3 and the eigenvalue condition of the mean
case are not nested, which explains why weak dependence at a quantile
is defined through A3. Section 3.3 gives primitive sufficient conditions
for the weak dependence assumption. For Markov-driven errors the rate
of convergence is $L_{NT}$, and for general $\beta$- and $\alpha$-mixing
errors it is $L_{NT}$ up to logarithmic factors. Section 3.4 derives
the limiting distribution and consistent estimators of the asymptotic
variance, and Section 3.5 shows that the rates are minimax optimal.
Figure 1 summarizes the logical structure of the results. The proof
of Theorem 3 is in Appendix A; the remaining proofs, including those
of Proposition 1 and Theorem 1, are in the Supplemental Appendix B.

\begin{figure}
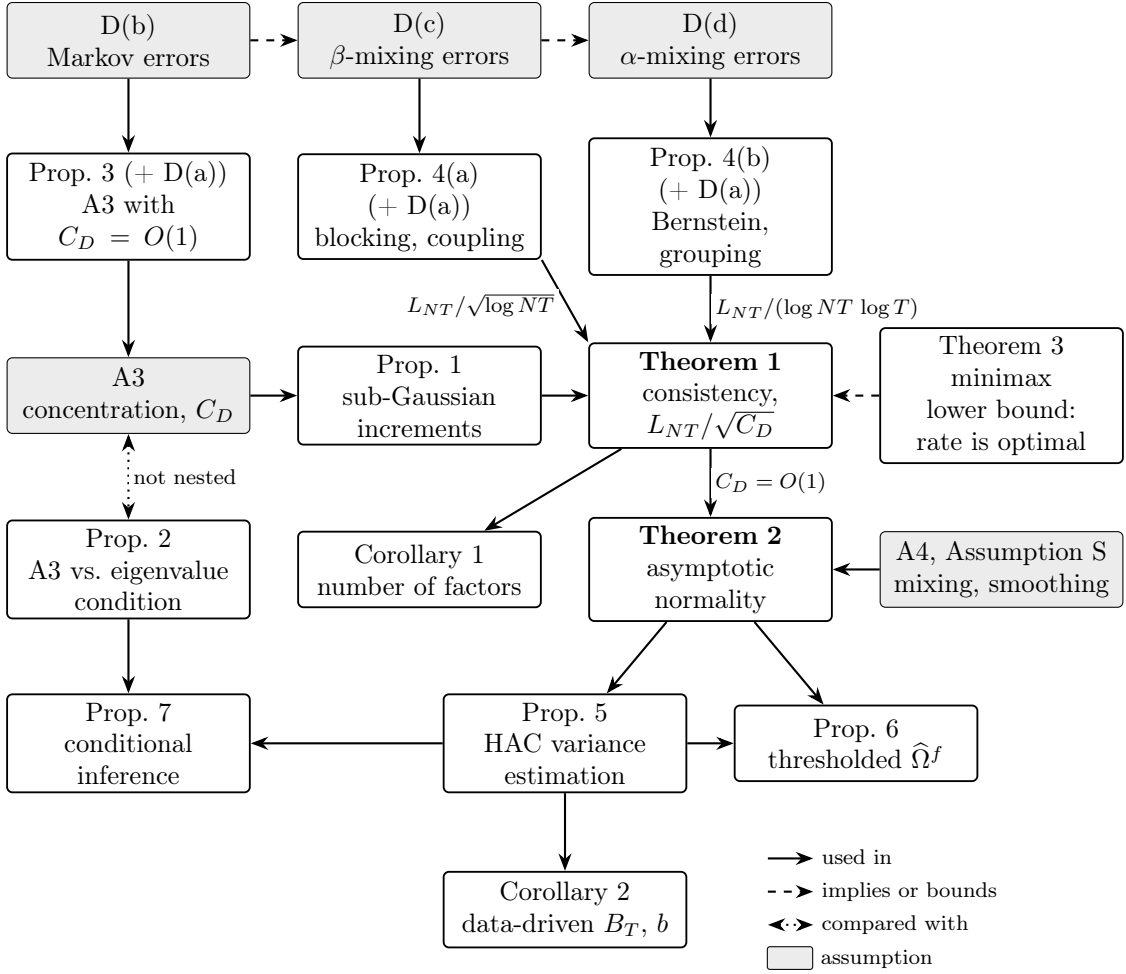

\begin{centering}
\roadmapfig
\par\end{centering}
\caption{\label{fig:roadmap}Structure of the results in Section 3. Shaded
boxes are assumptions. Solid arrows mean \textquotedbl is used in\textquotedbl ;
dashed arrows between assumptions mean \textquotedbl implies\textquotedbl ,
and the dashed arrow from Theorem 3 means \textquotedbl bounds from
below\textquotedbl . The dotted arrow means \textquotedbl is compared
with\textquotedbl : Proposition 2 shows that A3 and the eigenvalue
condition of the mean factor model are not nested. Rates on arrows
are those of the common component. All results maintain A1 and A2
(in Theorem 1, A2 conditional on $\mathcal{G}$).}
\end{figure}

\subsection{Assumptions and consistency}

Throughout, I write $\varepsilon_{it}=u_{it}(\tau)$ and $\theta=(\lambda_{1}',\dots,\lambda_{N}',f_{1}',\dots,f_{T}')'$,
and I let $\theta_{0}$ collect the true loadings and factors. Probabilities
and expectations are conditional on a $\sigma$-field $\mathcal{G}$
that contains the $\sigma$-field generated by $F_{0}$. Unless stated
otherwise, $\mathcal{G}=\sigma(F_{0})$, as in Assumptions A2 and
A3 below; Theorem 1 also allows larger $\sigma$-fields, which Section
3.2 uses to accommodate common volatility. I say that A2 or A3 holds
conditional on $\mathcal{G}$ if it holds with $F_{0}$ replaced by
$\mathcal{G}$. Since the bounds below do not depend on the conditioning
variables, they also hold unconditionally.

The analysis rests on three assumptions. Assumptions A1 and A2 are
standard conditions on the parameter space and the error densities
(CDG, Assumption 1). Assumption A3 formalizes the approximate condition
of Section 2.2.
\begin{description}
\item [{A1}] \textbf{(Parameter space)} For a constant $B<\infty$, $\theta\in\Theta^{r}$,
the set of parameters with $\Vert\lambda_{i}\Vert\le B$, $\Vert f_{t}\Vert\le B$,
$F'F/T=I_{r}$, and $\Lambda'\Lambda/N$ diagonal with nonincreasing
diagonal elements. The true parameter $\theta_{0}\in\Theta^{r}$ satisfies
$\Lambda_{0}'\Lambda_{0}/N=\mathrm{diag}(\sigma_{N1},\dots,\sigma_{Nr})\to\mathrm{diag}(\sigma_{1},\dots,\sigma_{r})$
with $\sigma_{1}>\dots>\sigma_{r}>0$, which is condition (P).
\item [{A2}] \textbf{(Density)} There exist constants $\kappa>0$ and $\underline{f}>0$,
not depending on $i,t,N,T$, such that, conditional on $F_{0}$, $\varepsilon_{it}$
has distribution function $F_{it}$ with $F_{it}(0)=\tau$ and density
$f_{it}$ satisfying $f_{it}(s)\ge\underline{f}$ for all $|s|\le\kappa$.
Thus $\kappa$ is the half-width of a fixed neighborhood of the true
quantile on which the error densities are bounded away from zero.
Together with $\underline{f}$ and $B$, it determines the curvature
constant $c_{0}$ in Lemma B.1 of the Supplemental Appendix.
\item [{A3}] \textbf{(Weak dependence)} There is a constant $C_{D}=C_{D,NT}$
such that, for all measurable functions $g_{it}:\mathbb{R}\to\mathbb{R}$
with oscillation $\sup g_{it}-\inf g_{it}\le c_{it}$ and all $x>0$,
\begin{equation}
P\Big(\Big|\sum_{i,t}\big[g_{it}(\varepsilon_{it})-\mathbb{E}[g_{it}(\varepsilon_{it})\mid F_{0}]\big]\Big|\ge x\ \Big|\ F_{0}\Big)\le2\exp\Big(-\frac{x^{2}}{C_{D}\sum_{i,t}c_{it}^{2}}\Big).\label{eq:A3}
\end{equation}
 Weak dependence means $C_{D}=O(1)$.
\end{description}
Assumption A3 contains the exact model as the special case $C_{D}=1/2$
and accommodates a broad range of dependence structures. Each of the
following known inequalities delivers A3 with an explicit constant,
provided that the stated dependence structure holds conditional on
$F_{0}$.
\begin{itemize}
\item \emph{Independence.} Hoeffding's inequality gives $C_{D}=1/2$.
\item \emph{Local dependence.} If the array $\{\varepsilon_{it}\}$ has
a dependency graph with fractional chromatic number $\chi^{*}$ and
maximum degree $\Delta$, then $C_{D}=\chi^{*}/2\le(\Delta+1)/2$
(Janson, 2004, Theorem 2.1). This case covers $m$-dependence in $t$
combined with bounded cross-sectional clusters.
\item \emph{Marton couplings.} Order the $NT$ errors as a sequence. If
the sequence admits a Marton coupling with mixing matrix $\Gamma$,
then $C_{D}=\Vert\Gamma\Vert^{2}/2$, where $\Vert\cdot\Vert$ is
the spectral norm (Paulin, 2015, Theorem 2.9). An important special
case arises when the errors are independent across $i$ and each unit's
series $\{\varepsilon_{it}\}_{t}$ is a uniformly ergodic Markov chain.
For unit $i$, let $d_{i}(t)=\max_{s}\sup_{x,y}d_{\mathrm{TV}}\big(\mathcal{L}(\varepsilon_{i,s+t}\mid\varepsilon_{is}=x),\mathcal{L}(\varepsilon_{i,s+t}\mid\varepsilon_{is}=y)\big)$,
where $\mathcal{L}(\cdot\mid\cdot)$ denotes a conditional law and
$d_{\mathrm{TV}}(P,Q)=\sup_{A}|P(A)-Q(A)|$ is the total variation
distance, with the supremum taken over Borel sets $A$. Let $t_{i}(\epsilon)=\min\{t:d_{i}(t)\le\epsilon\}$
be the mixing time at accuracy $\epsilon$, and define
\[
t_{\min,i}=\inf_{0\le\epsilon<1}t_{i}(\epsilon)\Big(\frac{2-\epsilon}{1-\epsilon}\Big)^{2}.
\]
 Paulin (2015, Corollary 2.11) bounds the moment generating function
of each unit's sum with constant $t_{\min,i}$. Because the units
are independent, multiplying these bounds gives $C_{D}=\max_{i}t_{\min,i}/2$,
which is $O(1)$ when the $t_{\min,i}$ are bounded uniformly in $i$.
For time-homogeneous chains, $t_{\min,i}\le9\,t_{\mathrm{mix},i}$,
where $t_{\mathrm{mix},i}$ is the usual mixing time to stationarity
at accuracy $1/4$ (Paulin, 2015, Definition 1.3 and Remark 2.12).
\end{itemize}
Not every concentration inequality for dependent data delivers A3.
The $\eta$-mixing inequality of Kontorovich and Ramanan (2008, Theorem
7.1) is stated with a single Lipschitz constant and therefore bounds
deviations in terms of $NT\max_{i,t}c_{it}^{2}$ rather than $\sum_{i,t}c_{it}^{2}$.
It does not deliver A3, which must hold with the heterogeneous weights
$c_{it}$ that arise in Proposition 1.

Assumption A3 enters the argument through a single inequality for
the centered criterion. To state it, let $\delta_{it}(\theta)=\lambda_{i}'f_{t}-\lambda_{0i}'f_{0t}$,
$X(\theta)=\Lambda F'/\sqrt{NT}$, and $d(\theta,\theta')=\Vert X(\theta)-X(\theta')\Vert_{F}$,
the root mean squared distance between two common components. Define
$M(\theta)=(NT)^{-1}\sum_{i,t}\rho_{\tau}(\varepsilon_{it}-\delta_{it}(\theta))$,
$\mathbb{M}(\theta)=\mathbb{E}[M(\theta)\mid\mathcal{G}]$, and $G(\theta)=\sum_{i,t}\{\rho_{\tau}(\varepsilon_{it}-\delta_{it}(\theta))-\rho_{\tau}(\varepsilon_{it})-\mathbb{E}[\rho_{\tau}(\varepsilon_{it}-\delta_{it}(\theta))-\rho_{\tau}(\varepsilon_{it})\mid\mathcal{G}]\}$.
Then $G(\theta_{0})=0$, and
\begin{equation}
NT\,[M(\theta)-M(\theta_{0})]=NT\,[\mathbb{M}(\theta)-\mathbb{M}(\theta_{0})]+G(\theta).\label{eq:decomp}
\end{equation}

\begin{prop}[Sub-Gaussian increments]
Suppose that Assumption A3 holds conditional on $\mathcal{G}$. Then,
for all $\theta,\theta'\in\Theta^{r}$ and $x>0$,
\begin{equation}
P\big(|G(\theta)-G(\theta')|\ge x\mid\mathcal{G}\big)\le2\exp\Big(-\frac{x^{2}}{4C_{D}\,NT\,d^{2}(\theta,\theta')}\Big).\label{eq:prop1}
\end{equation}
\end{prop}
Proposition 1 is the point at which idiosyncratic dependence enters
the consistency argument, and it is the reason for formulating the
approximate model through A3. Under independence, the corresponding
bound follows from Hoeffding's inequality (CDG, inequality (5)), and
CDG remark that their assumption could be relaxed if such a bound
continued to hold (their Remark 1.4). Assumption A3 is designed to
deliver this bound under dependence, with the price of dependence
summarized by the single constant $C_{D}$, and Section 3.3 identifies
the dependence structures under which it holds. The next result shows
that the increment bound is all that consistency requires of the dependence:
given the bound, the rate follows from the curvature of the population
criterion, whatever dependence produces it.
\begin{thm}[Consistency from concentration]
Suppose that Assumption A1 holds and that, for a $\sigma$-field
$\mathcal{G}$ that contains $\sigma(F_{0})$, Assumption A2 holds
conditional on $\mathcal{G}$ and the centered criterion satisfies
the increment bound $(\ref{eq:prop1})$ with a constant $C_{D}=C_{D,NT}$.
Let $(\widehat{\Lambda},\widehat{F})$ be a global minimizer of (\ref{eq:qfa})
 over $\Theta^{r}$ with the true $r=r(\tau)$, and let $\widehat{S}=\mathrm{diag}\big(\mathrm{sgn}(\widehat{f}_{k}'f_{0k}/T)\big)_{k\le r}$
collect the signs of the diagonal elements of $\widehat{F}'F_{0}/T$,
with the convention $\mathrm{sgn}(0)=1$. Then, as $N,T\to\infty$,
\[
\begin{gathered}\frac{1}{\sqrt{NT}}\Vert\widehat{\Lambda}\widehat{F}'-\Lambda_{0}F_{0}'\Vert=O_{P}\!\left(\widetilde{L}_{NT}^{-1}\right),\\
\frac{1}{\sqrt{N}}\bigl\Vert\widehat{\Lambda}-\Lambda_{0}\widehat{S}\bigr\Vert=O_{P}\!\left(\widetilde{L}_{NT}^{-1}\right),\qquad\frac{1}{\sqrt{T}}\bigl\Vert\widehat{F}-F_{0}\widehat{S}\bigr\Vert=O_{P}\!\left(\widetilde{L}_{NT}^{-1}\right),
\end{gathered}
\]
 where $\widetilde{L}_{NT}=L_{NT}/\sqrt{C_{D}}$, $L_{NT}=\min\{\sqrt{N},\sqrt{T}\}$,
and $\Vert\cdot\Vert$ is the Frobenius norm. The bound for the common
component holds uniformly over $\theta_{0}\in\Theta^{r}$, because
its constants depend on $\theta_{0}$ only through $B$, $\kappa$,
$\underline{f}$, and $C_{D}$, provided that $(\ref{eq:prop1})$
holds with the same $C_{D}$ for every $\theta_{0}$, as it does under
A3. In particular, under weak dependence ($C_{D}=O(1)$) the convergence
rate equals $L_{NT}$, the rate under independence, with dependence
affecting only the constant, and the estimated factor space is consistent
whenever $C_{D}=o(L_{NT}^{2})$.
\end{thm}
Theorem 1 separates the two ingredients of consistency. Assumption
A2 concerns the marginal distributions of the errors and determines
the curvature of the population criterion (Lemma B.1), whereas the
increment bound $(\ref{eq:prop1})$ is a property of their dependence
alone, which enters the rate only through $C_{D}$. By Proposition
1, A3 implies $(\ref{eq:prop1})$ with $\mathcal{G}=\sigma(F_{0})$,
so QFA attains the rate $L_{NT}/\sqrt{C_{D}}$ in the approximate
model, and Theorem 3 shows that this rate cannot be improved. This
is why A3 is formulated as a concentration property rather than as
a bound on the score covariances (Section 2.2); the covariances matter
only for the asymptotic variance (Section 3.4).

The proof of Theorem 1 combines the curvature bound with a low-rank
entropy bound (Lemma B.2), Dudley's chaining inequality (Lemma B.3)
and a peeling argument, and converts the rate for the common component
into rates for the factors and loadings by a singular-value perturbation
argument. Given the increment bound $(\ref{eq:prop1})$, the proof
uses the check function only through the curvature bound of Lemma
B.1 and the continuity of the criterion in $\theta$; the Lipschitz
property of $\rho_{\tau}$ enters only through Proposition 1, which
derives $(\ref{eq:prop1})$ from A3. The conclusion therefore holds
for any criterion of the form $(NT)^{-1}\sum_{i,t}\ell_{it}(\varepsilon_{it}-\delta_{it}(\theta))$
with continuous $\ell_{it}$ whose conditional mean has the curvature
of Lemma B.1 and whose centered version satisfies $(\ref{eq:prop1})$;
if each $\ell_{it}$ is $L$-Lipschitz, A3 delivers $(\ref{eq:prop1})$
with $C_{D}$ replaced by $L^{2}C_{D}$. Theorem 1 treats the number
of factors as known. The next result shows that the rank-minimization
and information-criterion estimators of $r(\tau)$ are consistent
under Assumption A3.
\begin{cor}[Number of factors]
Suppose that Assumptions A1--A3 hold. Let $k_{\max}>r$ and, as
in CDG (Section 3.2), let $\Theta^{l}$ be defined as $\Theta^{r}$
with $l$ factors. Assume that, for $l>r$, the true parameter is
admissible in $\Theta^{l}$, in the sense that the loadings $[\lambda_{0i}',0_{1\times(l-r)}]'$
and some completion of $F_{0}$ to $l$ orthonormal columns satisfy
the bound $B$. Let $\widehat{\theta}^{l}$ minimize (\ref{eq:qfa})
 over $\Theta^{l}$, and let $\widehat{d}_{1}\ge\dots\ge\widehat{d}_{k_{\max}}$
denote the diagonal elements of $(\widehat{\Lambda}^{k_{\max}})'\widehat{\Lambda}^{k_{\max}}/N$.
Consider the rank-minimization and information-criterion estimators
of CDG,
\[
\widehat{r}_{\mathrm{rank}}=\sum_{j=1}^{k_{\max}}\mathbb{I}\{\widehat{d}_{j}>P_{NT}\},\qquad\widehat{r}_{\mathrm{IC}}=\arg\min_{1\le l\le k_{\max}}\big\{ M(\widehat{\theta}^{l})+l\,P_{NT}\big\},
\]
 where $M(\theta)$ is the criterion in (\ref{eq:qfa})
. If
\begin{equation}
P_{NT}\to0\qquad\text{and}\qquad P_{NT}\,\widetilde{L}_{NT}^{2}=P_{NT}L_{NT}^{2}/C_{D}\to\infty,\label{eq:PNT}
\end{equation}
 then $P(\widehat{r}_{\mathrm{rank}}=r)\to1$ and $P(\widehat{r}_{\mathrm{IC}}=r)\to1$.
\end{cor}
Condition (\ref{eq:PNT})
 is the usual threshold condition (CDG, Theorems 2 and 3), namely
$P_{NT}\to0$ and $P_{NT}L_{NT}^{2}\to\infty$, with $L_{NT}$ replaced
by $\widetilde{L}_{NT}$. Under weak dependence ($C_{D}=O(1)$) the
two conditions coincide, so any threshold that is valid under independence
remains valid. More generally, the recommended threshold $P_{NT}=\widehat{d}_{1}L_{NT}^{-2/3}$
(CDG, equation (7)) satisfies (\ref{eq:PNT})
 whenever $C_{D}=o(L_{NT}^{4/3})$, because $\widehat{d}_{1}\to_{p}\sigma_{1}>0$.
For the information criterion, Corollary 1 also dispenses with the
upper bound on the error densities required in the independent case
(CDG, Theorem 3).
\begin{rem}
Theorem 1 concerns the global minimizer of (\ref{eq:qfa})
, whereas the IQR algorithm is guaranteed only to reach a stationary
point. This caveat applies equally to the existing theory under independence.
\end{rem}

\subsection{The concentration condition and the eigenvalue condition}

In the mean factor model, weak idiosyncratic dependence is formulated
as a bound on the largest eigenvalue of $\mathbb{E}[u_{t}u_{t}']$
(Chamberlain and Rothschild, 1983). The next result shows that this
condition and Assumption A3 are not nested, so that the mean-case
notion of an approximate factor model does not deliver the concentration
property on which the theory rests.
\begin{prop}[Eigenvalue condition and concentration]
(a) Let $W_{1},\dots,W_{T}$ be i.i.d. with $P(W_{t}=1)=P(W_{t}=-1)=1/2$,
and let $d\in(0,\min\{\tau,1-\tau\}/2)$. Conditional on $(W_{1},\dots,W_{T})$,
let the $\varepsilon_{it}$ be independent, and let $\varepsilon_{it}$
have mean zero, finite variance not depending on $W_{t}$, $P(\varepsilon_{it}\le0\mid W_{t})=\tau+dW_{t}$,
and a density that is bounded away from zero on $[-\kappa,\kappa]$
for some $\kappa>0$ not depending on $W_{t}$. Then $P(\varepsilon_{it}\le0)=\tau$,
A2 holds, and $\mathbb{E}[\varepsilon_{t}\varepsilon_{t}']=vI_{N}$
for a constant $v>0$, but $\mathrm{Cov}(\psi_{it},\psi_{jt})=d^{2}$
for all $i\neq j$, and every constant $C_{D}$ for which A3 holds
satisfies $C_{D}\ge d^{2}N/2$. If, in addition, $(W_{t})$ and the
$\varepsilon_{it}$ are independent of $F_{0}$, the $\varepsilon_{it}$
are identically distributed given $W_{t}$ with densities at least
$\underline{f}$ on $[-\kappa,\kappa]$, $d\le\underline{f}\kappa/2$,
and $\Lambda_{0}b=1_{N}$ for some $b\in\mathbb{R}^{r}$, with A1
holding for $\Lambda_{0}\widetilde{F}'$, where $\widetilde{F}=F_{0}+qb'$
and $q=(q(W_{1}),\dots,q(W_{T}))'$ collects the $\tau$-quantiles
of $\varepsilon_{it}$ given $W_{t}$, then $\Vert\widehat{\Lambda}\widehat{F}'-\Lambda_{0}F_{0}'\Vert/\sqrt{NT}\to_{p}(\mathbb{E}[q(W_{t})^{2}])^{1/2}>0$.\\
(b) Let $\tau=1/2$, let the $\zeta_{it}$ be i.i.d. $N(0,1)$, and
let $\xi$ be independent of them with $P(\xi=K)=P(\xi=-K)=p/2$ and
$P(\xi=0)=1-p$, where $p=e^{-2NT}/4$ and $K=e^{2NT}$. Let $\varepsilon_{it}=\zeta_{it}+\xi$.
Then A2 holds and A3 holds with $C_{D}=2$, but the largest eigenvalue
of $\mathbb{E}[\varepsilon_{t}\varepsilon_{t}']$ equals $1+NpK^{2}\to\infty$.\\
(c) Let the $\zeta_{it}$ be i.i.d. with mean zero, finite variance,
$P(\zeta_{it}\le0)=\tau$, and a density that is bounded away from
zero on $[-\kappa,\kappa]$ for some $\kappa>0$. Let $\sigma_{1},\dots,\sigma_{T}$
be i.i.d. with values in $[1,2]$ and a nondegenerate distribution,
where the laws of $\sigma_{t}$ and $\zeta_{it}$ do not depend on
$N$ or $T$, let $(\sigma_{t})$, $(\zeta_{it})$ and $F_{0}$ be
independent, and let $\varepsilon_{it}=\sigma_{t}\zeta_{it}$. Then
A2 holds and $\mathbb{E}[\varepsilon_{t}\varepsilon_{t}']=\mathbb{E}(\sigma_{t}^{2})\mathbb{E}(\zeta_{it}^{2})I_{N}$,
but every constant $C_{D}$ for which A3 holds satisfies $C_{D}\ge cN$
for a constant $c>0$ that does not depend on $N$ or $T$. With $\mathcal{G}=\sigma(F_{0},\sigma_{1},\dots,\sigma_{T})$,
however, $P(\varepsilon_{it}\le0\mid\mathcal{G})=\tau$, and A2 and
A3 hold conditional on $\mathcal{G}$, the latter with $C_{D}=1/2$.
\end{prop}
Part (a) is the important direction. The common variable $W_{t}$
shifts the $\tau$-quantile of every error in period $t$ in the same
direction, while leaving the conditional means, and hence all covariances
of the errors, unchanged. It is thus a pervasive quantile component
hidden in the idiosyncratic term, which second moments of the errors
cannot detect: the covariance matrix of the errors is diagonal, whereas
the covariance matrix of the scores has an eigenvalue of order $d^{2}N$.
Conditioning on $W_{t}$ does not help, because it would violate the
quantile restriction $P(\varepsilon_{it}\le0)=\tau$. Given $W_{t}$,
the $\tau$-quantile of $X_{it}$ is $\lambda_{0i}'f_{0t}+q(W_{t})$,
so QFA estimates a different common component, and the last claim
of (a) shows that it is inconsistent when the columns of $\Lambda_{0}$
span $1_{N}$, so that the factors absorb the shift. Otherwise the
shift acts as an additional factor that the eigenvalue condition does
not reveal. Assumption A3 excludes such components, and the eigenvalue
condition does not, which is why A3 is stated for bounded functions
of the errors rather than for their covariances.

Part (b) shows the converse. Because A3 restricts only bounded functions
of the errors, it is unaffected by tail events that dominate second
moments but have negligible probability, and by Theorem 1, QFA attains
the rate $L_{NT}$ in this model although the eigenvalue condition
fails.

The example is extreme by design. Independent errors without second
moments, such as the Cauchy errors in Table I of CDG, make a related
point: they satisfy A3 with $C_{D}=1/2$, whereas $\mathbb{E}[u_{t}u_{t}']$
is not defined.

Part (c) shows that A3 is stronger than consistency requires: it suffices
that A3 hold conditional on some $\sigma$-field $\mathcal{G}\supseteq\sigma(F_{0})$
under which A2, including the quantile restriction, still holds. This
matters because common volatility is a pervasive feature of macroeconomic
and financial panels. In (c), the scores are independent and the eigenvalue
condition holds, but the indicators $\mathbb{I}\{\varepsilon_{it}\le s\}$
at thresholds $s\neq0$ move together with $\sigma_{t}$, so A3 holds
only with $C_{D}\ge cN$. Conditioning on the volatility path restores
independence without violating the quantile restriction, because $\sigma_{t}$
rescales the errors but does not move their $\tau$-quantile. Under
A1, Theorem 1 then applies with $\mathcal{G}=\sigma(F_{0},\sigma_{1},\dots,\sigma_{T})$,
and QFA attains the rate $L_{NT}$.

The contrast between (a) and (c) identifies which common shocks matter
at a quantile. A common shock that moves the $\tau$-quantile of the
errors, such as $W_{t}$ in (a), acts like an omitted quantile factor:
A3 excludes it, and conditioning on it would violate the quantile
restriction, so Theorem 1 cannot accommodate it through the choice
of $\mathcal{G}$. A common shock that only rescales the errors, such
as $\sigma_{t}$ in (c), is harmless at $\tau$ once the analysis
conditions on it.

\subsection{Sufficient conditions for weak dependence}

Assumption A3 is a high-level condition. This subsection provides
primitive conditions on the error processes under which the conclusions
of Theorem 1 hold, either at the rate $L_{NT}$ or with a logarithmic
loss. The conditions are expressed in terms of mixing coefficients.
For unit $i$, let
\[
\beta_{i}(k)=\sup_{s}\,\beta\big(\sigma(\varepsilon_{iu}:u\le s),\,\sigma(\varepsilon_{iu}:u\ge s+k)\big),
\]
 where, for $\sigma$-fields $\mathcal{A}$ and $\mathcal{B}$, $\beta(\mathcal{A},\mathcal{B})=\frac{1}{2}\sup\sum_{a,b}|P(A_{a}\cap B_{b})-P(A_{a})P(B_{b})|$,
with the supremum taken over finite partitions $\{A_{a}\}\subset\mathcal{A}$
and $\{B_{b}\}\subset\mathcal{B}$. The strong mixing coefficient
$\alpha_{i}(k)$ is defined analogously with $\alpha(\mathcal{A},\mathcal{B})=\sup_{A\in\mathcal{A},B\in\mathcal{B}}|P(A\cap B)-P(A)P(B)|$.
All coefficients are computed under the conditional law given $F_{0}$,
and the constants below do not depend on $N$ or $T$.
\begin{description}
\item [{D}] \textbf{(Dependence structure)} (a) \emph{Block independence.}
The units can be partitioned into $H$ classes such that, within each
class, the series $\{\varepsilon_{it}\}_{t}$ are mutually independent
conditional on $F_{0}$. (b) \emph{Markov-driven errors.} Conditional
on $F_{0}$, $\varepsilon_{it}=\varphi_{it}(Z_{it})$ for measurable
functions $\varphi_{it}$. For each $i$, $\{Z_{it}\}_{t}$ is a time-homogeneous
Markov chain on a countably generated state space that is started
from its stationary distribution and is irreducible, aperiodic, and
geometrically ergodic. The transition kernels of these chains belong
to a finite set. (c) \emph{Absolute regularity.} $\beta_{i}(k)\le C_{\beta}e^{-c_{\beta}k}$
for all $i$ and $k\ge1$. (d) \emph{Strong mixing.} $\alpha_{i}(k)\le e^{-2c_{\alpha}k}$
for all $i$ and $k\ge1$.
\end{description}
Part (a) accommodates cross-sectional independence ($H=1$) as well
as finite-range cross-sectional dependence, under which units farther
apart than a fixed distance are independent. Parts (b)--(d) are nested.
A stationary, geometrically ergodic Markov chain is geometrically
$\beta$-mixing, so (b) implies (c). Because $\alpha\le\beta$ and
$\alpha\le1/4$, (c) implies (d) for sufficiently small $c_{\alpha}$.
I begin with the strongest condition, D(b).
\begin{prop}[Markov-driven errors]
Suppose that Assumptions A1 and A2 and conditions D(a) and D(b) hold.
Then Assumption A3 holds with $C_{D}=O(1)$, and the conclusions of
Theorem 1 and Corollary 1 hold with the rate $L_{NT}$.
\end{prop}
The functions $\varphi_{it}$ may depend on $t$, which accommodates
scale factors generated by the factor structure. Condition D(b) covers
the error models most commonly used in econometrics:
\begin{itemize}
\item stationary causal ARMA errors with coprime polynomials and an everywhere
positive innovation density, which are functions of a geometrically
ergodic state-space chain (Mokkadem, 1988);
\item GARCH(1,1) errors under the conditions of Carrasco and Chen (2002),
which include an innovation density that is positive near zero and
a negative Lyapunov exponent; here the chain is $Z_{it}=(\sigma_{it}^{2},z_{it})$;
\item regime-switching errors $\varepsilon_{it}=\mu_{S_{it}}+s_{S_{it}}\eta_{it}$
with i.i.d. $\eta_{it}$ and an irreducible, aperiodic, finite-state
regime chain $S_{it}$. Markov-switching autoregressions require additional
stability conditions.
\end{itemize}
The uniformity of the constant across units stems from the finiteness
of the set of kernels in D(b). This requirement excludes, for example,
autoregressive coefficients that vary continuously across units. The
leading example with a single kernel is the design that CDG use to
study dependent errors.
\begin{example}[CDG's Monte Carlo design]
CDG (Section 5.2) simulate
\[
X_{it}=\lambda_{1i}f_{1t}+\lambda_{2i}f_{2t}+(\lambda_{3i}f_{3t})\,e_{it},\qquad e_{it}=\phi e_{i,t-1}+w_{it},\qquad w_{it}=v_{it}+\gamma\sum_{j\in\mathcal{N}_{J}(i)}v_{jt},
\]
 where $\mathcal{N}_{J}(i)=\{j\in\{1,\dots,N\}:0<|j-i|\le J\}$, the
$v_{it}$ are i.i.d. $N(0,1)$ and independent of the factors and
loadings, $|\phi|<1$, $J$ is fixed, $\lambda_{3i}f_{3t}>0$, and
$e_{it}$ is the stationary solution. I denote by $\phi$ and $\gamma$
the coefficients that CDG call $\beta$ and $\rho$, to avoid confusion
with $\rho_{\tau}$ and with the common use of $\beta$ for loadings.
Supplemental Appendix B shows that D(a) holds with $H=2J+1$ and that
D(b) holds with a single kernel. By Proposition 3, the errors therefore
satisfy Assumption A3 with $C_{D}=4K_{1}M_{\phi}(2J+1)$, where $K_{1}$
is an absolute constant and $M_{\phi}$ depends only on $\phi$. CDG's
designs M1 and M3 are special cases. Design M2 has independent $t_{3}$
errors, for which A3 holds with $C_{D}=1/2$. Theorem 1 therefore
covers the evidence of CDG that QFA performs well under mildly dependent
errors. One caveat is unrelated to dependence: the Gaussian factors
in this design are unbounded, so A1 and A2 hold only on events with
probability approaching one, as is also the case for the theory of
CDG.
\end{example}
When the errors are mixing but not Markov-driven, Assumption A3 need
not hold with a bounded constant. The next result shows that the conclusions
of Theorem 1 continue to hold, with a logarithmic loss in the rate.
\begin{prop}[Mixing errors]
Suppose that Assumptions A1 and A2 and condition D(a) hold. (a) Under
D(c), let $q_{NT}=\lceil2\log(NT)/c_{\beta}\rceil$. Then the conclusions
of Theorem 1 and Corollary 1 hold with $\widetilde{L}_{NT}=L_{NT}/\sqrt{Hq_{NT}}\asymp L_{NT}/\sqrt{\log(NT)}$.
(b) Under D(d), the bounds of Theorem 1 for the common component,
for $\widehat{\Lambda}$, and for $\widehat{F}$ hold with the rate
\[
\frac{1}{\sqrt{NT}}\Vert\widehat{\Lambda}\widehat{F}'-\Lambda_{0}F_{0}'\Vert=O_{P}\Big(\frac{(\log NT)(\log T)}{L_{NT}}\Big),
\]
 and the conclusions of Corollary 1 hold if $P_{NT}\to0$ and $P_{NT}L_{NT}^{2}\to\infty$
faster than $(\log NT)^{2}(\log T)^{2}$.
\end{prop}
Part (a) couples the errors, on an event with probability approaching
one, with independent blocks of length $q_{NT}$ (Berbee, 1979, Corollary
4.2.5) and applies Hoeffding's inequality to the block sums. This
yields Proposition 1 with $C_{D}=Hq_{NT}$. Under strong mixing, no
exact coupling is available, and couplings that control the distance
between a block and its copy have constants that grow with the block
dimension. Part (b) therefore applies the Bernstein inequality of
Merlevède, Peligrad and Rio (2009) separately to groups of summands
of comparable magnitude and then takes a union bound over a net. The
logarithmic factors arise from these general arguments. By Proposition
3, they are absent for Markov-driven errors, and Section 3.5 shows
that the rate $L_{NT}$ cannot be improved.

\subsection{Asymptotic distribution}

Inference on the factors and loadings requires their limiting distribution.
Because the check function is not differentiable, I derive it for
a smoothed version of QFA with a differentiable criterion, as in CDG.
Let $k$ be a kernel supported on $[-1,1]$, let $K(u)=\int_{u}^{1}k(v)\,dv$,
let $h=h_{NT}$ be a bandwidth, and let $\ell_{h}(u)=[\tau-K(u/h)]u$
be the smoothed check function. Let $(\widetilde{\Lambda},\widetilde{F})$
minimize $(NT)^{-1}\sum_{i,t}\ell_{h}(x_{it}-\lambda_{i}'f_{t})$
over $\Theta^{r}$, and let $\widetilde{S}=\mathrm{diag}\big(\mathrm{sgn}(\widetilde{f}_{k}'f_{0k}/T)\big)_{k\le r}$.
I impose the smoothing and regularity conditions of CDG (their Assumption
2), which I refer to as Assumption S: $k$ is a symmetric, twice continuously
differentiable kernel of order $m\ge8$; the error densities are $m+2$
times continuously differentiable with bounded derivatives and bounded
away from zero on compact sets; the true loadings and factors are
interior points, which, because $N$ and $T$ grow, I take to hold
with a fixed margin, $\Vert\lambda_{0i}\Vert\le B-\epsilon_{0}$ and
$\Vert f_{0t}\Vert\le B-\epsilon_{0}$ for all $i$ and $t$ and some
$\epsilon_{0}>0$; $N\propto T$ and $h\propto T^{-c}$ with $1/m<c<1/6$;
and the matrices
\[
\Phi_{i}=\lim_{T}\frac{1}{T}\sum_{t=1}^{T}f_{it}(0)f_{0t}f_{0t}',\qquad\Psi_{t}=\lim_{N}\frac{1}{N}\sum_{i=1}^{N}f_{it}(0)\lambda_{0i}\lambda_{0i}'
\]
 are positive definite. Let $\psi_{it}=\tau-\mathbb{I}\{u_{it}\le0\}$.
Under dependence, the distribution theory requires the following additional
conditions, all stated conditional on $F_{0}$.
\begin{description}
\item [{A4}] \textbf{(Dependence for the distribution theory)} (a) \emph{Serial
absolute regularity.} The error vector $u_{t}=(u_{1t},\dots,u_{Nt})'$
has $\beta$-mixing coefficients $\beta(k)=\sup_{t}\beta\big(\sigma(u_{s}:s\le t),\sigma(u_{s}:s\ge t+k)\big)\le C_{A}N^{\varsigma}a^{k}$
for constants $C_{A}<\infty$, $\varsigma\ge0$, and $a\in(0,1)$.
(b) \emph{Finite-range cross-sectional dependence.} For a fixed $J^{*}$
and some ordering of the units, the families $\{u_{it}:i\in S,t\le T\}$
and $\{u_{jt}:j\in S',t\le T\}$ are independent whenever $|i-j|>J^{*}$
for all $i\in S$ and $j\in S'$. (c) \emph{Bivariate densities.}
The pairs $(u_{is},u_{it})$, $s\neq t$, and $(u_{it},u_{jt})$,
$i\neq j$, have joint densities bounded by a constant that does not
depend on $i,j,s,t,N,T$. (d) \emph{Long-run variances.} The limits
\[
\Omega_{i}^{\lambda}=\lim_{T}\frac{1}{T}\sum_{s,t=1}^{T}\mathbb{E}[\psi_{is}\psi_{it}]\,f_{0s}f_{0t}',\qquad\Omega_{t}^{f}=\lim_{N}\frac{1}{N}\sum_{i,j=1}^{N}\mathbb{E}[\psi_{it}\psi_{jt}]\,\lambda_{0i}\lambda_{0j}'
\]
 exist and are positive definite, and
\[
\frac{1}{\sqrt{T}}\sum_{t=1}^{T}\psi_{it}f_{0t}\to_{d}N(0,\Omega_{i}^{\lambda}),\qquad\frac{1}{\sqrt{N}}\sum_{i=1}^{N}\psi_{it}\lambda_{0i}\to_{d}N(0,\Omega_{t}^{f}).
\]
\end{description}
In the expansion of the smoothed estimator, Assumption A3 controls
the terms that are linear in the errors, and A4 controls the quadratic
terms. Under independence the off-diagonal quadratic terms have mean
zero, whereas under dependence they do not, and A4 ensures that they
are negligible.

Part (d) is a high-level condition in the spirit of Bai (2003). If,
in addition, each unit's own $\beta$-mixing coefficients are bounded
by $Ca^{k}$, then, because $\psi_{it}$ is bounded, (d) can be obtained
from a central limit theorem for strongly mixing triangular arrays
with bounded weights $f_{0t}$, and from a central limit theorem for
finite-range dependent arrays with weights $\lambda_{0i}$, once the
limits are positive definite and the variance-regularity and Lindeberg-type
conditions of the chosen theorem are verified for the given weights.
Because the factors may form a nonstationary sequence, these conditions
do not follow from mixing alone, and I therefore keep (d) as a high-level
condition.

The factor $N^{\varsigma}$ in (a) allows the mixing coefficients
of the whole cross-section to grow with $N$. They do so for the error
process of Example 1, which is a Gaussian VAR(1) with diagonal autoregressive
matrix $\phi I_{N}$. In that design, each unit's series depends only
on the innovations of units within distance $J$, so (b) holds with
$J^{*}=2J$.
\begin{thm}[Asymptotic distribution]
Suppose that Assumptions A1--A3 hold with $C_{D}=O(1)$, that Assumption
A4 holds, and that Assumption S holds. Then, for each $i$ and $t$,
\[
\sqrt{T}\,(\widetilde{\lambda}_{i}-\widetilde{S}\lambda_{0i})\to_{d}N\big(0,\Phi_{i}^{-1}\Omega_{i}^{\lambda}\Phi_{i}^{-1}\big),\qquad\sqrt{N}\,(\widetilde{f}_{t}-\widetilde{S}f_{0t})\to_{d}N\big(0,\Psi_{t}^{-1}\Omega_{t}^{f}\Psi_{t}^{-1}\big).
\]
\end{thm}
Under independence across $i$ and $t$, $\mathbb{E}[\psi_{is}\psi_{it}]=\tau(1-\tau)\mathbb{I}\{s=t\}$
and $\mathbb{E}[\psi_{it}\psi_{jt}]=\tau(1-\tau)\mathbb{I}\{i=j\}$.
Hence $\Omega_{i}^{\lambda}=\tau(1-\tau)I_{r}$ by the normalization,
and $\Omega_{t}^{f}=\tau(1-\tau)\Sigma_{\Lambda}$ with $\Sigma_{\Lambda}=\lim_{N}\Lambda_{0}'\Lambda_{0}/N=\mathrm{diag}(\sigma_{1},\dots,\sigma_{r})$,
neither of which depends on its index, so Theorem 4 of CDG is the
special case of Theorem 2 under independence. Under weak dependence,
the rates and the Hessians $\Phi_{i}$ and $\Psi_{t}$ are unchanged,
and only the middle matrices differ: $\Omega_{i}^{\lambda}$ is a
long-run variance over time, and $\Omega_{t}^{f}$ is a cross-sectional
long-run variance. The proof in Supplemental Appendix B shows that
the additional terms generated by dependence are $O_{P}(N^{-1})$
and $O_{P}(\log(NT)/T)$ and are therefore negligible. Consequently,
the estimators remain free of incidental-parameter bias. An important
implication for inference is that the covariance estimators of CDG
(their Remark 4.4), which target $\tau(1-\tau)\Phi_{i}^{-2}$ and
$\tau(1-\tau)\Psi_{t}^{-1}\Sigma_{\Lambda}\Psi_{t}^{-1}$, are inconsistent
under dependence. Proposition 5 provides consistent alternatives.

These alternatives combine heteroskedasticity and autocorrelation
consistent (HAC) estimators of the middle matrices with kernel estimators
of the Hessians. Let $\widehat{\psi}_{it}=\tau-\mathbb{I}\{x_{it}\le\widetilde{\lambda}_{i}'\widetilde{f}_{t}\}$,
let $w$ be an even kernel with $|w|\le1$, $w(0)=1$, $w$ continuous
at $0$, and $w(x)=0$ for $|x|>1$ (for example, the Bartlett or
Parzen kernel), and let $B_{T}$ be a lag truncation parameter. Define
\[
\widehat{\Omega}_{i}^{\lambda}=\sum_{|l|<T}w(l/B_{T})\,\widehat{\Gamma}_{i}(l),\qquad\widehat{\Gamma}_{i}(l)=\frac{1}{T}\sum_{t=1}^{T-l}\widehat{\psi}_{it}\widehat{\psi}_{i,t+l}\widetilde{f}_{t}\widetilde{f}_{t+l}'\ (l\ge0),\quad\widehat{\Gamma}_{i}(-l)=\widehat{\Gamma}_{i}(l)',
\]
 
\[
\widehat{\Omega}_{t}^{f}=\frac{1}{N}\sum_{i,j:\,|i-j|\le J^{*}}\widehat{\psi}_{it}\widehat{\psi}_{jt}\widetilde{\lambda}_{i}\widetilde{\lambda}_{j}',
\]
 and let $\widetilde{\Phi}_{i}=(Tb)^{-1}\sum_{t}l(\widetilde{u}_{it}/b)\widetilde{f}_{t}\widetilde{f}_{t}'$
and $\widetilde{\Psi}_{t}=(Nb)^{-1}\sum_{i}l(\widetilde{u}_{it}/b)\widetilde{\lambda}_{i}\widetilde{\lambda}_{i}'$
be the kernel estimators of the Hessians in CDG (their Remark 4.4),
where $\widetilde{u}_{it}=x_{it}-\widetilde{\lambda}_{i}'\widetilde{f}_{t}$,
$l$ is a kernel, and $b$ is a bandwidth. Set $\widehat{V}_{i}^{\lambda}=\widetilde{\Phi}_{i}^{-1}\widehat{\Omega}_{i}^{\lambda}\widetilde{\Phi}_{i}^{-1}$
and $\widehat{V}_{t}^{f}=\widetilde{\Psi}_{t}^{-1}\widehat{\Omega}_{t}^{f}\widetilde{\Psi}_{t}^{-1}$.
\begin{prop}[Variance estimation]
Suppose that the conditions of Theorem 2 hold with $\widetilde{S}=I_{r}$
and, in addition, that (i) each unit's own strong mixing coefficients
satisfy $\alpha_{i}(k)\le Ca^{k}$, where $C$ and $a\in(0,1)$ do
not depend on $N$ or $T$; (ii) the ordering of the units and $J^{*}$
in A4(b) are known; (iii) $B_{T}\to\infty$ and $B_{T}=o(L_{NT}^{2/3})$;
and (iv) $l$ is a continuous kernel supported on $[-1,1]$ whose
first two derivatives exist and are bounded on $(-1,1)$, $b\to0$,
and $Tb^{3}\to\infty$. Then $\widehat{V}_{i}^{\lambda}\to_{p}\Phi_{i}^{-1}\Omega_{i}^{\lambda}\Phi_{i}^{-1}$
and $\widehat{V}_{t}^{f}\to_{p}\Psi_{t}^{-1}\Omega_{t}^{f}\Psi_{t}^{-1}$.
\end{prop}
Because $\widetilde{\lambda}_{i}'\widetilde{f}_{t}$ is invariant
to the signs in $\widetilde{S}$, the normalization $\widetilde{S}=I_{r}$
entails no loss of generality. Condition (i) strengthens A4(a) at
the level of an individual unit and ensures that the autocovariances
of $\psi_{it}f_{0t}$ are summable uniformly in $N$. Condition (ii)
is needed for the cross-sectional long-run variance: as in the mean
case, $\Omega_{t}^{f}$ cannot be estimated without some known structure
in the cross-section. The upper bound on $B_{T}$ controls both the
sampling variability of the autocovariances and the effect of the
estimation error in $\widehat{\psi}_{it}$, which enters through indicator
functions. In finite samples, the banded estimator $\widehat{\Omega}_{t}^{f}$
need not be positive definite. Because the probability that it is
positive definite tends to one, replacing it on the complementary
event by any positive definite matrix, such as $\tau(1-\tau)\widetilde{\Lambda}'\widetilde{\Lambda}/N$,
does not affect Proposition 5.

Proposition 5 uses the ordering in A4(b) to estimate $\Omega_{t}^{f}$.
This requirement can be dispensed with when the cross-sectional covariances
of the score are stable over time. In that case, the covariances $\sigma_{ij}=\mathbb{E}[\psi_{it}\psi_{jt}\mid F_{0}]$
can be estimated by time averages and then thresholded, in the spirit
of Bickel and Levina (2008). Because $\psi_{it}$ depends on $u_{it}$
only through its sign, $\sigma_{ij}$ is determined by the copula
of $(u_{it},u_{jt})$ and does not depend on the scales of the errors.
Let $\widehat{\sigma}_{ij}=T^{-1}\sum_{t}\widehat{\psi}_{it}\widehat{\psi}_{jt}$,
let $\omega_{NT}$ be a threshold, and define
\[
\widehat{\Omega}^{f}=\frac{1}{N}\sum_{i,j=1}^{N}\widehat{\sigma}_{ij}\,\mathbb{I}\{|\widehat{\sigma}_{ij}|>\omega_{NT}\}\,\widetilde{\lambda}_{i}\widetilde{\lambda}_{j}'.
\]

\begin{prop}[Estimation without a known ordering]
Suppose that the conditions of Theorem 2 hold with $\widetilde{S}=I_{r}$,
where the ordering in A4(b) exists but need not be known, and that
$l$ and $b$ satisfy condition (iv) of Proposition 5. Suppose, in
addition, that (i) $\sigma_{ij}$ does not depend on $t$, and (ii)
for every pair $(i,j)$, the series $\{(u_{it},u_{jt})\}_{t}$ is
strongly mixing with coefficients at most $e^{-2c_{\alpha}k}$ for
a constant $c_{\alpha}>0$ that does not depend on $i,j,N,T$. Let
$r_{NT}=\sqrt{\log(NT)}/(hL_{NT})$ and $\omega_{NT}=\ell_{NT}\big(r_{NT}+\sqrt{\log N/T}\big)$,
with $\ell_{NT}\to\infty$ and $\omega_{NT}\to0$. Then $\widehat{\Omega}^{f}\to_{p}\Omega^{f}$,
the common value of $\Omega_{t}^{f}$, and $\widetilde{\Psi}_{t}^{-1}\widehat{\Omega}^{f}\widetilde{\Psi}_{t}^{-1}\to_{p}\Psi_{t}^{-1}\Omega^{f}\Psi_{t}^{-1}$.
\end{prop}
By A4(b), each row of $(\sigma_{ij})$ has at most $2J^{*}+1$ nonzero
entries. With probability approaching one, thresholding sets every
zero entry to zero without knowledge of the ordering, and this suffices
for consistency. Exact recovery of the nonzero entries is not needed
and would require them to exceed the threshold.

The threshold depends on the uniform rate $r_{NT}$ of the smoothed
estimators, $\max_{i}\Vert\widetilde{\lambda}_{i}-\lambda_{0i}\Vert+\max_{t}\Vert\widetilde{f}_{t}-f_{0t}\Vert=O_{P}(r_{NT})$,
which is established in Lemma B.7. Under Assumption S, $r_{NT}\to0$
because $h\propto T^{-c}$ with $c<1/6$ and $L_{NT}\propto T^{1/2}$.
Since $r_{NT}$ is known up to a constant, the threshold is feasible.

Proposition 6 thus removes the need to know the ordering, but not
the finite-range structure itself: A4(b) must hold for some ordering,
so that each row of $(\sigma_{ij})$ is sparse. Proposition 5 exploits
a known finite-range structure, Proposition 6 an unknown one, and
neither covers dense cross-sectional dependence. Extending Proposition
6 to approximately sparse matrices, as in Bickel and Levina (2008),
would require a uniform bound on $\sum_{j}|\sigma_{ij}|^{q}$ for
some $q\in[0,1)$ in place of A4(b), and I do not pursue it here.

Propositions 5 and 6 take the tuning parameters $B_{T}$ and $b$
as given. The next result shows that standard data-dependent choices
satisfy the conditions of Proposition 5.
\begin{cor}[Data-dependent tuning parameters]
(a) Let $B_{T}^{*}$ and $b^{*}$ be deterministic sequences that
satisfy conditions (iii) and (iv) of Proposition 5. If $P(B_{T}^{*}/2\le\widehat{B}_{T}\le2B_{T}^{*})\to1$
and $\widehat{b}/b^{*}\to_{p}1$, then the conclusions of Proposition
5 hold with $\widehat{B}_{T}$ and $\widehat{b}$ in place of $B_{T}$
and $b$. (b) Suppose that $N\propto T$. With the Parzen kernel,
the plug-in rule of Andrews (1991), $\widehat{B}_{T}=2.6614\,(\widehat{\alpha}(2)T)^{1/5}$,
satisfies the conditions in (a) whenever $\widehat{\alpha}(2)\to_{p}\alpha^{*}\in(0,\infty)$.
With the Epanechnikov kernel $l(z)=0.75(1-z^{2})\mathbb{I}\{|z|\le1\}$,
the rule of thumb $\widehat{b}=2.34\,\widehat{s}\,T^{-1/5}$, where
$\widehat{s}$ is the sample standard deviation of the residuals $\widetilde{u}_{it}$,
satisfies the conditions in (a) whenever $\widehat{s}\to_{p}s^{*}\in(0,\infty)$.
\end{cor}
Here $\widehat{\alpha}(2)$ is computed, as in Andrews (1991), from
univariate AR(1) fits to the elements of $\widehat{\psi}_{it}\widetilde{f}_{t}$,
and $\alpha^{*}(2)$ denotes its probability limit; the rule of thumb
is that of Silverman (1986, Section 3.4.2), adapted to the Epanechnikov
kernel.

The Bartlett kernel combined with the rule of Andrews yields $\widehat{B}_{T}\propto T^{1/3}$,
which violates the sufficient condition (iii) of Proposition 5. I
therefore use the Parzen kernel, although a sharper bound on the sampling
error may accommodate the Bartlett kernel as well.

The plug-in rule requires $\alpha^{*}(2)>0$, which fails when the
scores are serially uncorrelated, as under independence. In that case,
a lower bound such as $\max\{\widehat{B}_{T},T^{1/5}\}$ keeps the
corollary applicable.

Similarly, the Hall--Sheather bandwidth $b\propto T^{-1/3}$, which
is popular in quantile regression, violates $Tb^{3}\to\infty$ in
condition (iv). These exclusions concern sufficient conditions only;
they do not show that the Bartlett kernel or the Hall--Sheather bandwidth
leads to inconsistent variance estimators.

In finite samples, $\widehat{\alpha}(2)$ is nonnegative by construction.
I truncate the estimated autoregressive coefficients to $[-0.97,0.97]$,
so that it is finite, and use $\max\{\widehat{B}_{T},T^{1/5}\}$,
which is well defined even when $\widehat{\alpha}(2)=0$.

The results so far condition on $F_{0}$. Proposition 2(c) shows that
common volatility is accommodated by conditioning on a larger $\sigma$-field,
and the same device extends the distribution theory.
\begin{prop}[Inference conditional on common shocks]
Let $\mathcal{G}\supseteq\sigma(F_{0})$ be a $\sigma$-field such
that $P(\varepsilon_{it}\le0\mid\mathcal{G})=\tau$ for all $i$ and
$t$, and suppose that the conditions of Theorem 2, or of any of Propositions
5 and 6 and Corollary 2, hold conditional on $\mathcal{G}$. Here
$f_{it}$ is the density of $\varepsilon_{it}$ conditional on $\mathcal{G}$;
the moments and limits in A4, Assumption S and the definitions of
$\Phi_{i}$, $\Psi_{t}$, $\Omega_{i}^{\lambda}$ and $\Omega_{t}^{f}$
are conditional on $\mathcal{G}$ and are assumed to exist almost
surely; and the deterministic sequences and limits in Corollary 2
may be $\mathcal{G}$-measurable. Then the conclusions of that result
hold conditional on $\mathcal{G}$. In particular, with $\widetilde{S}=I_{r}$,
$(\widehat{V}_{i}^{\lambda})^{-1/2}\sqrt{T}(\widetilde{\lambda}_{i}-\lambda_{0i})$
and $(\widehat{V}_{t}^{f})^{-1/2}\sqrt{N}(\widetilde{f}_{t}-f_{0t})$,
or the latter with $\widetilde{\Psi}_{t}^{-1}\widehat{\Omega}^{f}\widetilde{\Psi}_{t}^{-1}$
in place of $\widehat{V}_{t}^{f}$ under Proposition 6, converge in
distribution to $N(0,I_{r})$, both conditional on $\mathcal{G}$
and unconditionally.
\end{prop}
Unconditionally, the limit of $\sqrt{N}(\widetilde{f}_{t}-f_{0t})$
is a mixture of normals with a $\mathcal{G}$-measurable variance,
but the studentized estimators remain pivotal. Consider the common-volatility
model of Proposition 2(c), or more generally $\varepsilon_{it}=\sigma_{t}\zeta_{it}$
with $(\zeta_{it})$ identically distributed with density $f_{\zeta}$,
independent of $(\sigma_{t})$, and satisfying A3 with $C_{D}=O(1)$,
A4 and Assumption S, and with $\sigma_{t}$ bounded and bounded away
from zero. The scores depend only on the $\zeta_{it}$, whereas the
density of $\varepsilon_{it}$ at zero given $\mathcal{G}$ is $f_{\zeta}(0)/\sigma_{t}$.
Hence $\Psi_{t}=f_{\zeta}(0)\sigma_{t}^{-1}\Sigma_{\Lambda}$, and
the asymptotic variance of $\widetilde{f}_{t}$ is proportional to
$\sigma_{t}^{2}$, so its unconditional limit is a scale mixture over
the law of $\sigma_{t}$. The estimators of Propositions 5 and 6 reflect
this without modification, because $\widetilde{\Psi}_{t}$ is computed
from the cross-section at date $t$ and the score covariances do not
depend on volatility. Section 4.5 confirms this in simulations.

The volatility path may be deterministic, driven by its own shocks,
as in stochastic-volatility or regime-switching models, or driven
by past factors, but not by past idiosyncratic errors, because conditioning
on the path would then reveal information about those errors; such
feedback is weak in large panels, where idiosyncratic errors largely
average out across units. Unit-specific GARCH errors are covered by
A3 directly (Proposition 3).

\subsection{Optimality of the rates}

Sections 3.1 and 3.3 establish the rate $L_{NT}$ under A3 with $C_{D}=O(1)$
(Theorem 1 and Proposition 3), the rate $L_{NT}/\sqrt{\log(NT)}$
under $\beta$-mixing (Proposition 4(a)), and the rate $L_{NT}/\{(\log NT)(\log T)\}$
under $\alpha$-mixing (Proposition 4(b)). More generally, Theorem
1 gives the rate $L_{NT}/\sqrt{C_{D}}$ for any dependence constant.
The following result shows that this rate cannot be improved: no estimator
attains a faster rate uniformly over models that satisfy A1--A3 with
dependence constant $C_{D}$.
\begin{thm}[Minimax lower bound]
Let $r=1$, $B\ge3$ and $s>0$, let $c_{s}=\min\{1,32/(s^{2}\log2)\}$,
and let $m=m_{NT}$ be an integer with $1\le m\le c_{s}L_{NT}^{2}$.
Let the variables $z$ below be i.i.d. $N(0,1)$, and let $u_{it}=s(z_{\lceil i/m\rceil,t}-\Phi^{-1}(\tau))$
if $N\le T$ and $u_{it}=s(z_{i,\lceil t/m\rceil}-\Phi^{-1}(\tau))$
if $N>T$, so that the errors are identical within blocks of $m$
consecutive units (if $N\le T$) or periods (if $N>T$) and independent
across blocks. Then Assumptions A1--A3 hold with $C_{D}=m/2$, and
there exist a constant $c>0$ depending only on $s$ and an absolute
constant $c'>0$ such that, for all sufficiently large $N$ and $T$,
\[
\inf_{\widehat{C}}\sup_{C_{0}}P_{C_{0}}\Big(\frac{1}{\sqrt{NT}}\Vert\widehat{C}-C_{0}\Vert\ge c\,\frac{\sqrt{C_{D}}}{L_{NT}}\Big)\ge c'.
\]
 Here the supremum is taken over common components $C_{0}=\Lambda_{0}F_{0}'$
with $(\Lambda_{0},F_{0})\in\Theta^{1}$ and $\Lambda_{0}'\Lambda_{0}/N\ge1/4$,
which replaces the limit condition (P) for fixed $N$ and $T$, and
the infimum is taken over all estimators.
\end{thm}
With $m=1$ the errors are independent and $C_{D}=1/2$, so the rate
$L_{NT}$ of Theorem 1 and Proposition 3 is minimax optimal. The model
with $m=1$ also satisfies the assumptions of Proposition 4 with $H=1$,
so the rates there are optimal up to the factors $\sqrt{\log(NT)}$
and $(\log NT)(\log T)$.

For general $m$, the lower bound matches the bound $\sqrt{C_{D}}/L_{NT}$
of Theorem 1, which holds uniformly over the same class. In the worst
case over models with dependence constant $C_{D}$, the constant therefore
enters the rate exactly through $\sqrt{C_{D}}$: weak dependence costs
only a constant, and stronger dependence costs precisely the factor
that Theorem 1 attributes to it.

When $m$ is of order $L_{NT}^{2}$, the lower bound is bounded away
from zero, so no estimator of the common component is uniformly consistent,
and the condition $C_{D}=o(L_{NT}^{2})$ in Theorem 1 cannot be relaxed
uniformly over models with dependence constant $C_{D}$.

Because
\[
\frac{1}{\sqrt{NT}}\Vert\widehat{\Lambda}\widehat{F}'-\Lambda_{0}F_{0}'\Vert\le B\left(\frac{1}{\sqrt{N}}\Vert\widehat{\Lambda}-\Lambda_{0}\widehat{S}\Vert+\frac{1}{\sqrt{T}}\Vert\widehat{F}-F_{0}\widehat{S}\Vert\right),
\]
the lower bound also applies, for estimators in $\Theta^{1}$, to
the larger of the normalized loading and factor errors, so that no
estimator can make both smaller than order $\sqrt{C_{D}}/L_{NT}$.
It does not give separate minimax bounds for the loadings and for
the factors.

Whether the logarithmic factors in Proposition 4 are necessary for
mixing errors that are not Markov-driven is an open probabilistic
question. For Markov chains, Dedecker and Gouëzel (2015, Theorem 2
and Remark 4) show that the sub-Gaussian inequality holds if and only
if the chain is geometrically ergodic; to my knowledge, the general
case is open.

\section{Monte Carlo}

The experiments test the predictions of Section 3: how the precision
of QFA changes as dependence strengthens (Theorem 1), when the selection
of the number of factors breaks down (Corollary 1), whether the results
extend to the error processes of Propositions 3 and 4 and to tail
quantiles, and whether the inference procedures of Section 3.4 deliver
correct coverage (Theorem 2, Propositions 5--7, and Corollary 2).
The designs build on those of CDG, which satisfy the assumptions of
Theorem 1 by Example 1, and let the dependence become stronger and
more varied than in their experiments.

\subsection{Design and implementation}

The consistency experiments use the factor structure of CDG (their
equation (10)), so that the results under independence can be compared
with theirs, 
\[
X_{it}=\lambda_{1i}f_{1t}+\lambda_{2i}f_{2t}+(\lambda_{3i}f_{3t})\,e_{it},
\]
 where $f_{1t}=0.8f_{1,t-1}+\epsilon_{1t}$, $f_{2t}=0.5f_{2,t-1}+\epsilon_{2t}$
and $f_{3t}=|g_{t}|$, the variables $\lambda_{1i}$, $\lambda_{2i}$,
$\epsilon_{1t}$, $\epsilon_{2t}$ and $g_{t}$ are i.i.d. $N(0,1)$,
and $\lambda_{3i}$ is i.i.d. $U[1,2]$.

The errors follow Example 1: $e_{it}=\phi e_{i,t-1}+w_{it}$ with
$w_{it}=v_{it}+\gamma\sum_{0<|j-i|\le J}v_{jt}$, where $v_{it}$
is i.i.d. $N(0,1)$ and $J=3$. The autoregressive coefficient $\phi$
governs the serial correlation of $e_{it}$, and the parameters $\gamma$
and $J$ govern its cross-sectional correlation: $e_{it}$ is correlated
with the errors of the units within distance $2J$ of unit $i$, and
these correlations become stronger as $\gamma$ increases. Larger
values of $\phi$ or $\gamma$ therefore mean stronger dependence.
Designs M1 ($\phi=\gamma=0$) and M3 ($\phi=\gamma=0.2$) are those
of CDG. In the other designs, which I label by $(\phi,\gamma)$, I
rescale $e_{it}$ to unit variance, so that differences across designs
reflect the strength of dependence rather than the scale of the errors.

The grid extends $\phi$ to 0.95 but stops $\gamma$ at 0.5 because
the two parameters differ in how much dependence they can generate.
After rescaling, the sum of the correlations of $e_{it}$ with $e_{jt}$
over all $j$ equals $(1+2J\gamma)^{2}/(1+2J\gamma^{2})$, which is
at most $2J+1=7$ and already equals 6.4 at $\gamma=0.5$, whereas
the sum of the autocorrelations of $e_{it}$ over all lags equals
$(1+\phi)/(1-\phi)$, which is 3, 9 and 39 at $\phi=0.5$, 0.8 and
0.95 and is unbounded as $\phi\to1$. Stronger cross-sectional dependence
would therefore require a larger range $J$ rather than a larger $\gamma$.

At $\tau\neq0.5$ the quantile factor model has $r(\tau)=3$ factors
with loadings $(\lambda_{1i},\lambda_{2i},q_{\tau}\lambda_{3i})$,
where $q_{\tau}$ is the $\tau$-quantile of $e_{it}$. At the median
$q_{0.5}=0$, so $f_{3t}$ drops out and $r(0.5)=2$. I do not report
$\tau=0.75$. In all my designs the loadings $\lambda_{1i}$ and $\lambda_{2i}$,
the factor innovations and the errors are symmetric about zero, so
$-X$ has the same distribution as $X$; since $\rho_{\tau}(u)=\rho_{1-\tau}(-u)$,
the results at $1-\tau$ have the same distribution as those at $\tau$,
as the nearly identical $\tau=0.25$ and $\tau=0.75$ columns of CDG's
Table II confirm.

I compute QFA by the IQR algorithm, started at the principal components
estimator. Each quantile regression is solved by the MM algorithm
of Hunter and Lange (2000), applied to all units (or all periods)
at once because they share the same regressors, and the iterations
stop when the relative change in the estimated common component falls
below $10^{-3}$. The number of factors is estimated by the rank-minimization
estimator with $k=8$ and the threshold $P_{NT}=\widehat{d}_{1}L_{NT}^{-2/3}$
(CDG, equation (7)). For each design I report the average of $\widehat{r}$,
the average adjusted $R^{2}$ from the regression of each true factor
on the $\widehat{r}$ estimated factors, and the root mean squared
error of the common component, $\mathrm{RMSE}=\Vert\widehat{\Lambda}\widehat{F}'-\Lambda_{0}F_{0}'\Vert/\sqrt{NT}$,
computed with the true $r(\tau)$. The results are based on 1,000
replications per design.

\subsection{Strength of dependence}

Table 1 reports the average $\widehat{r}$ and the recovery of the
factors for $\tau\in\{0.25,0.5\}$ and $N=T\in\{50,100,200\}$. In
each design the dependence constant $C_{D}$ does not change with
$N$ and $T$, so Theorem 1 and Corollary 1 predict that $\widehat{r}$
converges to $r(\tau)$ and that the estimated factors span the true
ones, whatever the value of $(\phi,\gamma)$. Dependence should therefore
matter only in finite samples, through $C_{D}$: at a given sample
size, stronger dependence should lower the $R^{2}$ and move $\widehat{r}$
away from $r(\tau)$, and the gap should narrow as $N$ and $T$ grow.
I expect the effect to be largest for the scale factor $f_{3t}$,
which is the hardest factor to recover even under independence, and
for $\phi=0.95$, for which $C_{D}$ is not small relative to the
threshold condition $C_{D}=o(L_{NT}^{4/3})$ of Corollary 1 at the
sample sizes considered ($L_{NT}^{4/3}\approx34$ at $N=T=200$).

Three features stand out. First, for M1 and M3 my results match Table
II of CDG closely at $N=T=200$, apart from a lower $R^{2}$ for the
scale factor under M1 (0.89 against 0.93); in smaller samples they
are lower for the weak scale factor at $\tau=0.25$, most likely because
of implementation details of the IQR algorithm.

Second, up to $\phi=0.8$ and $\gamma=0.5$, dependence leaves the
recovery of the factors essentially unchanged at $N=T=200$: the average
$\widehat{r}$ is between 2.93 and 3.00 at $\tau=0.25$ and between
2.00 and 2.02 at the median, and the $R^{2}$ for $f_{1t}$ and $f_{2t}$
is at least 0.98. Cross-sectional dependence mainly affects the scale
factor $f_{3t}$, whose $R^{2}$ at $\tau=0.25$ falls from between
0.89 and 0.91 when $\gamma=0$ to between 0.72 and 0.74 when $\gamma=0.5$.

Third, with $\phi=0.95$ the rank estimator tends to overestimate
$r(\tau)$, particularly at the median, where it often detects a spurious
factor that is correlated with $f_{3t}$. Overall, Table 1 bears out
the predictions: moderate dependence slows the recovery of the scale
factor but leaves the number of factors and the other two factors
essentially unaffected at $N=T=200$, and only the most persistent
errors disturb the selection of $r(\tau)$.

\begin{table}[htbp]
\caption{\label{tab:mc-dep}Number of factors and recovery of the factors under
increasing dependence}

\begin{centering}
{\footnotesize{}%
\begin{tabular}{lccccccccc}
\hline 
 &  & \multicolumn{4}{c}{$\tau=0.25$, $r=3$} & \multicolumn{4}{c}{$\tau=0.5$, $r=2$}\tabularnewline
\cline{3-10}
Design & $N=T$ & $\widehat{r}$ & $f_{1t}$ & $f_{2t}$ & $f_{3t}$ & $\widehat{r}$ & $f_{1t}$ & $f_{2t}$ & $f_{3t}$\tabularnewline
\hline 
M1 & 50 & 2.01 & 0.954 & 0.805 & 0.111 & 1.90 & 0.963 & 0.811 & 0.017\tabularnewline
 & 100 & 2.54 & 0.983 & 0.958 & 0.474 & 1.99 & 0.986 & 0.961 & 0.001\tabularnewline
 & 200 & 2.96 & 0.993 & 0.986 & 0.893 & 2.00 & 0.994 & 0.987 & 0.000\tabularnewline
\hline 
M3 & 50 & 2.34 & 0.936 & 0.788 & 0.211 & 2.19 & 0.955 & 0.821 & 0.040\tabularnewline
 & 100 & 2.78 & 0.979 & 0.943 & 0.553 & 2.08 & 0.982 & 0.951 & 0.009\tabularnewline
 & 200 & 3.00 & 0.990 & 0.982 & 0.858 & 2.00 & 0.992 & 0.984 & 0.000\tabularnewline
\hline 
(0, 0.5) & 50 & 2.14 & 0.946 & 0.777 & 0.110 & 2.08 & 0.961 & 0.797 & 0.025\tabularnewline
 & 100 & 2.52 & 0.982 & 0.950 & 0.311 & 2.06 & 0.985 & 0.952 & 0.008\tabularnewline
 & 200 & 2.94 & 0.993 & 0.985 & 0.741 & 2.00 & 0.993 & 0.987 & 0.000\tabularnewline
\hline 
(0.5, 0) & 50 & 2.09 & 0.952 & 0.815 & 0.166 & 1.91 & 0.964 & 0.811 & 0.018\tabularnewline
 & 100 & 2.52 & 0.984 & 0.944 & 0.475 & 1.99 & 0.986 & 0.954 & 0.001\tabularnewline
 & 200 & 2.97 & 0.993 & 0.986 & 0.914 & 2.00 & 0.994 & 0.988 & 0.001\tabularnewline
\hline 
(0.5, 0.5) & 50 & 2.19 & 0.948 & 0.775 & 0.146 & 2.15 & 0.962 & 0.820 & 0.038\tabularnewline
 & 100 & 2.50 & 0.983 & 0.942 & 0.317 & 2.05 & 0.985 & 0.943 & 0.007\tabularnewline
 & 200 & 2.93 & 0.993 & 0.985 & 0.739 & 2.00 & 0.994 & 0.987 & 0.001\tabularnewline
\hline 
(0.8, 0) & 50 & 2.15 & 0.952 & 0.783 & 0.254 & 2.06 & 0.965 & 0.806 & 0.085\tabularnewline
 & 100 & 2.48 & 0.984 & 0.939 & 0.434 & 2.05 & 0.987 & 0.955 & 0.016\tabularnewline
 & 200 & 2.95 & 0.993 & 0.986 & 0.893 & 2.00 & 0.994 & 0.988 & 0.001\tabularnewline
\hline 
(0.8, 0.5) & 50 & 2.27 & 0.948 & 0.760 & 0.221 & 2.19 & 0.962 & 0.791 & 0.099\tabularnewline
 & 100 & 2.59 & 0.982 & 0.942 & 0.362 & 2.24 & 0.986 & 0.954 & 0.043\tabularnewline
 & 200 & 2.93 & 0.993 & 0.986 & 0.724 & 2.02 & 0.993 & 0.986 & 0.002\tabularnewline
\hline 
(0.95, 0) & 50 & 2.45 & 0.954 & 0.769 & 0.608 & 2.44 & 0.967 & 0.807 & 0.513\tabularnewline
 & 100 & 2.93 & 0.987 & 0.940 & 0.711 & 2.79 & 0.989 & 0.953 & 0.464\tabularnewline
 & 200 & 3.14 & 0.994 & 0.987 & 0.863 & 2.58 & 0.994 & 0.988 & 0.240\tabularnewline
\hline 
(0.95, 0.5) & 50 & 2.40 & 0.955 & 0.729 & 0.495 & 2.40 & 0.960 & 0.786 & 0.423\tabularnewline
 & 100 & 2.94 & 0.987 & 0.940 & 0.605 & 2.84 & 0.990 & 0.946 & 0.405\tabularnewline
 & 200 & 3.21 & 0.994 & 0.986 & 0.749 & 2.84 & 0.995 & 0.989 & 0.251\tabularnewline
\hline 
\end{tabular}}{\footnotesize\par}
\par\end{centering}
{\footnotesize\emph{Notes:}}{\footnotesize{} Averages over 1,000 replications.
The columns $\widehat{r}$ report the average of the rank-minimization
estimator; the columns $f_{1t}$, $f_{2t}$ and $f_{3t}$ report the
average adjusted $R^{2}$ from the regression of each true factor
on the $\widehat{r}$ estimated factors (negative values are reported
as zero). M1 and M3 are the designs of CDG; the other designs are
labeled by $(\phi,\gamma)$ and have unit-variance errors. The coefficient
$\phi$ sets the serial correlation of the errors and $\gamma$ (with
$J=3$) their cross-sectional correlation; larger values of either
mean stronger dependence.}{\footnotesize\par}
\end{table}

Table 2 reports the RMSE of the common component in Panel A and the
scaled error $L_{NT}\cdot\mathrm{RMSE}$ in Panel B. By Theorem 1,
the scaled error is bounded in probability, with a bound proportional
to $\sqrt{C_{D}}$. The theorem therefore makes two predictions: in
each design the scaled error should be stable or decreasing as $N=T$
grows, because the rate $L_{NT}$ does not depend on dependence, and
its level should increase with the strength of dependence, because
the constant does. The first prediction is asymptotic and need not
hold at my sample sizes when $C_{D}$ is large, as for $\phi=0.95$.

Both predictions hold for $\phi\le0.8$: in all these designs the
scaled error in Panel B decreases with the sample size, and at $N=T=200$
and $\tau=0.25$ it rises from 4.7 under independence to 7.7 when
$(\phi,\gamma)=(0.8,0.5)$. Dependence therefore lowers the precision
of QFA gradually and through the constant rather than the rate, as
Theorem 1 predicts. With $\phi=0.95$ and $\tau=0.25$, the RMSE in
Panel A still falls with the sample size, from 1.31 at $N=T=50$ to
0.77 at $N=T=200$ when $\gamma=0$. The rate $L_{NT}$ implies that
the RMSE should halve when $N=T$ quadruples, but here it falls only
to 59 percent of its initial value, so the scaled error rises. The
bound of Theorem 1 is proportional to $\sqrt{C_{D}}/L_{NT}$, and
it is informative only when $\sqrt{C_{D}}$ is small relative to $L_{NT}$.
For such persistent errors this is not yet the case: $C_{D}$ grows
with the long-run variance factor $(1+\phi)/(1-\phi)$, which equals
39, and $\sqrt{39}\approx6.2$ is comparable to $L_{NT}$, which lies
between 7.1 and 14.1 over my sample sizes.

\begin{table}[htbp]
\caption{\label{tab:mc-rate}Estimation error of the common component}

\begin{centering}
{\footnotesize{}%
\begin{tabular}{lcccccc}
\hline 
 & \multicolumn{3}{c}{$\tau=0.25$} & \multicolumn{3}{c}{$\tau=0.5$}\tabularnewline
\cline{2-7}
Design & $50$ & $100$ & $200$ & $50$ & $100$ & $200$\tabularnewline
\hline 
\multicolumn{7}{l}{\emph{Panel A: RMSE}}\tabularnewline
M1 & 0.839 & 0.504 & 0.330 & 0.409 & 0.276 & 0.192\tabularnewline
M3 & 1.021 & 0.642 & 0.441 & 0.515 & 0.327 & 0.222\tabularnewline
(0, 0.5) & 1.006 & 0.626 & 0.429 & 0.468 & 0.292 & 0.195\tabularnewline
(0.5, 0) & 0.869 & 0.546 & 0.369 & 0.446 & 0.295 & 0.202\tabularnewline
(0.5, 0.5) & 1.049 & 0.667 & 0.461 & 0.512 & 0.307 & 0.205\tabularnewline
(0.8, 0) & 1.007 & 0.686 & 0.471 & 0.528 & 0.327 & 0.219\tabularnewline
(0.8, 0.5) & 1.146 & 0.783 & 0.543 & 0.637 & 0.348 & 0.224\tabularnewline
(0.95, 0) & 1.305 & 1.033 & 0.772 & 0.956 & 0.457 & 0.261\tabularnewline
(0.95, 0.5) & 1.332 & 1.078 & 0.803 & 1.008 & 0.518 & 0.263\tabularnewline
\hline 
\multicolumn{7}{l}{\emph{Panel B: $L_{NT}\cdot\mathrm{RMSE}$}}\tabularnewline
M1 & 5.93 & 5.04 & 4.67 & 2.89 & 2.76 & 2.71\tabularnewline
M3 & 7.22 & 6.42 & 6.24 & 3.64 & 3.27 & 3.13\tabularnewline
(0, 0.5) & 7.12 & 6.26 & 6.07 & 3.31 & 2.92 & 2.76\tabularnewline
(0.5, 0) & 6.15 & 5.46 & 5.22 & 3.15 & 2.95 & 2.86\tabularnewline
(0.5, 0.5) & 7.41 & 6.67 & 6.52 & 3.62 & 3.07 & 2.90\tabularnewline
(0.8, 0) & 7.12 & 6.86 & 6.66 & 3.73 & 3.27 & 3.10\tabularnewline
(0.8, 0.5) & 8.10 & 7.83 & 7.68 & 4.50 & 3.48 & 3.17\tabularnewline
(0.95, 0) & 9.23 & 10.33 & 10.92 & 6.76 & 4.57 & 3.69\tabularnewline
(0.95, 0.5) & 9.42 & 10.78 & 11.35 & 7.13 & 5.18 & 3.71\tabularnewline
\hline 
\end{tabular}}{\footnotesize\par}
\par\end{centering}
{\footnotesize\emph{Notes:}}{\footnotesize{} $\mathrm{RMSE}=\Vert\widehat{\Lambda}\widehat{F}'-\Lambda_{0}F_{0}'\Vert/\sqrt{NT}$
with the true number of factors, averaged over 1,000 replications.
Panel A reports the RMSE and Panel B the RMSE multiplied by $L_{NT}=\sqrt{N}$,
which by Theorem 1 is bounded in probability. Column headings give
$N=T$. Designs as in Table 1.}{\footnotesize\par}
\end{table}

\subsection{Near-unit-root errors}

Corollary 1 requires $C_{D}=o(L_{NT}^{4/3})$ for the threshold of
Section 4.1, and Theorem 1 requires $C_{D}=o(L_{NT}^{2})$ for consistency,
a requirement that Theorem 3 shows cannot be relaxed in general. To
approach these boundaries, I let the autoregressive coefficient tend
to one with the sample size, $\phi_{T}=1-T^{-a}$ with $a\in\{0.25,0.5,0.75,1\}$,
and set $\gamma=0$ and $\tau=0.25$. For uniformly ergodic Markov
chains the dependence constant is proportional to the mixing time
(Section 3.1), and the mixing time of an AR(1) chain is of order $(1-\phi_{T})^{-1}=T^{a}$.
Treating $C_{D}$ as proportional to $T^{a}$, and with $N=T$, the
heuristic boundaries are therefore $a=2/3$ for the threshold and
$a=1$ for consistency. I therefore expect $\widehat{r}$ to overestimate
$r(0.25)$ when $a>2/3$, and the RMSE to keep falling with the sample
size when $a<1$ but not when $a=1$.

Table 3 confirms both boundaries. The RMSE of the common component
falls with the sample size for $a=0.25$ and $a=0.5$, falls slowly
for $a=0.75$, and does not fall for $a=1$, where it remains near
1.45. For $a=0.75$ the rank estimator overestimates the number of
factors at $N=T=200$, with an average $\widehat{r}$ of 3.44 and
a frequency of correct selection of 0.54, as the threshold condition
of Corollary 1 predicts. The $R^{2}$ for the factors, by contrast,
rises toward one in all designs. Because the errors are independent
across units, the cross-sectional information that identifies the
factors is unaffected by serial dependence, and the loss of precision
is concentrated in the loadings, which are estimated from time series
of length $T$. For $a=1$ the errors are nearly constant over the
sample, so the idiosyncratic term is close to $\lambda_{3i}(e_{i}-q_{\tau})f_{3t}$
for a unit-specific constant $e_{i}$. This term has the same structure
as the third common component, which explains why $\widehat{r}$ is
again close to three while the common component is estimated inconsistently.

\begin{table}
\caption{\label{tab:mc-unit}Near-unit-root errors, $\phi_{T}=1-T^{-a}$}

\begin{centering}
{\footnotesize{}%
\begin{tabular}{lcccccccc}
\hline 
$a$ & $N=T$ & $\phi_{T}$ & $\widehat{r}$ & $P(\widehat{r}=r)$ & $f_{1t}$ & $f_{2t}$ & $f_{3t}$ & RMSE\tabularnewline
\hline 
0.25 & 50 & 0.624 & 2.09 & 0.16 & 0.956 & 0.805 & 0.174 & 0.898\tabularnewline
 & 100 & 0.684 & 2.50 & 0.49 & 0.984 & 0.944 & 0.451 & 0.601\tabularnewline
 & 200 & 0.734 & 2.97 & 0.97 & 0.993 & 0.986 & 0.914 & 0.432\tabularnewline
\hline 
0.5 & 50 & 0.859 & 2.23 & 0.27 & 0.949 & 0.788 & 0.310 & 1.084\tabularnewline
 & 100 & 0.900 & 2.62 & 0.46 & 0.984 & 0.937 & 0.513 & 0.842\tabularnewline
 & 200 & 0.929 & 3.02 & 0.80 & 0.993 & 0.987 & 0.854 & 0.676\tabularnewline
\hline 
0.75 & 50 & 0.947 & 2.36 & 0.41 & 0.949 & 0.749 & 0.541 & 1.282\tabularnewline
 & 100 & 0.968 & 3.01 & 0.67 & 0.988 & 0.934 & 0.835 & 1.173\tabularnewline
 & 200 & 0.981 & 3.44 & 0.54 & 0.996 & 0.989 & 0.949 & 1.076\tabularnewline
\hline 
1 & 50 & 0.980 & 2.46 & 0.54 & 0.949 & 0.740 & 0.752 & 1.486\tabularnewline
 & 100 & 0.990 & 2.92 & 0.92 & 0.994 & 0.942 & 0.951 & 1.458\tabularnewline
 & 200 & 0.995 & 3.03 & 0.97 & 0.998 & 0.996 & 0.989 & 1.439\tabularnewline
\hline 
\end{tabular}}{\footnotesize\par}
\par\end{centering}
{\footnotesize\emph{Notes:}}{\footnotesize{} $\tau=0.25$, $r=3$, $\gamma=0$,
unit-variance errors, 1,000 replications. Columns $f_{1t}$, $f_{2t}$
and $f_{3t}$ report average adjusted $R^{2}$, as in Table 1; RMSE
is the root mean squared error of the common component.}{\footnotesize\par}
\end{table}

\subsection{Other error processes and tail quantiles}

Table 4 replaces the errors of Example 1 with three processes, all
scaled to unit variance: GARCH(1,1) errors and Markov regime-switching
errors, both covered by Proposition 3, and autoregressive errors whose
coefficients $\phi_{i}\sim U[0,0.9]$ vary continuously across units.
The last design violates the finite set of kernels in D(b) but satisfies
D(c), so it is covered by Proposition 4(a) rather than Proposition
3.

Combined with Theorem 1, these propositions predict that the RMSE
is of order $1/L_{NT}$, which equals $N^{-1/2}$ when $N=T$. Each
time $N=T$ doubles, the RMSE should therefore be multiplied by about
$1/\sqrt{2}\approx0.71$. In the heterogeneous autoregressive design,
Proposition 4 allows a slightly slower decline, because its rate contains
an additional logarithmic factor. The propositions also predict that
$\widehat{r}$ is consistent.

The results agree with these predictions. In all three designs the
RMSE is multiplied by a factor between 0.63 and 0.70 each time $N=T$
doubles, close to the predicted 0.71, and at $N=T=200$ the frequency
of correct selection of $r(\tau)$ is 1.00 at the median and 0.94
or higher at $\tau=0.25$, except for regime-switching errors at $\tau=0.25$,
for which it is 0.67. In that design the third loading $q_{0.25}\lambda_{3i}$
is small, because the mixture of scales concentrates probability near
zero, and $f_{3t}$ is harder to detect. The rate of Theorem 1 therefore
extends to all three processes. There is no sign of the logarithmic
loss of Proposition 4 in the heterogeneous autoregressive design,
which suggests that this loss may reflect the generality of the proof
rather than the behavior of the estimator.

\begin{table}
\caption{\label{tab:mc-other}Other error processes}

\begin{centering}
{\footnotesize{}%
\begin{tabular}{lccccccccc}
\hline 
 &  & \multicolumn{4}{c}{$\tau=0.25$, $r=3$} & \multicolumn{4}{c}{$\tau=0.5$, $r=2$}\tabularnewline
\cline{3-10}
Errors & $N=T$ & $\widehat{r}$ & $P(\widehat{r}=r)$ & $f_{3t}$ & RMSE & $\widehat{r}$ & $P(\widehat{r}=r)$ & $f_{2t}$ & RMSE\tabularnewline
\hline 
GARCH(1,1) & 50 & 1.99 & 0.09 & 0.096 & 0.847 & 1.93 & 0.76 & 0.813 & 0.396\tabularnewline
 & 100 & 2.48 & 0.48 & 0.422 & 0.531 & 1.99 & 0.97 & 0.956 & 0.263\tabularnewline
 & 200 & 2.94 & 0.94 & 0.876 & 0.345 & 2.00 & 1.00 & 0.989 & 0.181\tabularnewline
\hline 
Regime switching & 50 & 1.98 & 0.08 & 0.067 & 0.820 & 1.90 & 0.77 & 0.831 & 0.309\tabularnewline
 & 100 & 2.27 & 0.29 & 0.228 & 0.529 & 1.98 & 0.97 & 0.967 & 0.196\tabularnewline
 & 200 & 2.67 & 0.67 & 0.599 & 0.338 & 2.00 & 1.00 & 0.994 & 0.133\tabularnewline
\hline 
Heterogeneous AR & 50 & 2.16 & 0.26 & 0.202 & 0.926 & 1.98 & 0.72 & 0.809 & 0.461\tabularnewline
 & 100 & 2.52 & 0.50 & 0.406 & 0.609 & 1.99 & 0.95 & 0.945 & 0.302\tabularnewline
 & 200 & 2.97 & 0.97 & 0.849 & 0.424 & 2.00 & 1.00 & 0.988 & 0.204\tabularnewline
\hline 
\end{tabular}}{\footnotesize\par}
\par\end{centering}
{\footnotesize\emph{Notes:}}{\footnotesize{} 1,000 replications; unit-variance
errors. GARCH(1,1): $e_{it}=\sigma_{it}z_{it}$, $\sigma_{it}^{2}=0.05+0.10e_{i,t-1}^{2}+0.85\sigma_{i,t-1}^{2}$.
Regime switching: $e_{it}=s_{S_{it}}\eta_{it}$ with scales proportional
to $(0.5,1.5)$ and a two-state chain $S_{it}$ that stays in its
current state with probability 0.95. Heterogeneous AR: $\phi_{i}\sim U[0,0.9]$,
$\gamma=0.2$, $J=3$. Innovations are Gaussian and, except in the
heterogeneous AR design, independent across units. Column $P(\widehat{r}=r)$
reports the frequency of correct selection of the number of factors.
Column $f_{3t}$ ($f_{2t}$) reports the average adjusted $R^{2}$
of the factor with the lowest $R^{2}$ at that quantile; RMSE as in
Table 2.}{\footnotesize\par}
\end{table}

Table 5 reports results for the tail quantiles $\tau\in\{0.05,0.10\}$.
The theory applies to any fixed $\tau\in(0,1)$, so the rate is the
same in the tails, but two constants change in opposite directions.
The loading $q_{\tau}\lambda_{3i}$ of the scale factor grows with
$|q_{\tau}|$, which should make $f_{3t}$ easier to detect, while
the density of the errors at the quantile falls, which should make
the common component harder to estimate.

The density matters because Assumption A2 bounds it from below by
$\underline{f}$ near the quantile, and $\underline{f}$ determines
the curvature constant $c_{0}$ of Lemma B.1, which enters the constant
in Theorem 1: a lower density means a flatter objective function and
a less precise estimator.

Both effects appear: $\widehat{r}$ is close to three already at $N=T=100$,
and the RMSE is larger than at $\tau=0.25$. Under independence and
at $N=T=200$, for example, the RMSE is 0.48 at $\tau=0.05$ and 0.40
at $\tau=0.10$, against 0.33 at $\tau=0.25$ (Table 2, Panel A).

The cost of dependence is similar to that at the center of the distribution:
with $(\phi,\gamma)=(0.5,0.5)$ the RMSE at $N=T=200$ is about 30
and 36 percent higher than under independence at $\tau=0.05$ and
$\tau=0.10$, against 40 percent at $\tau=0.25$ in Table 2. The results
of Section 4.2 therefore carry over to the tails.

\begin{table}
\caption{\label{tab:mc-tails}Tail quantiles}

\begin{centering}
{\footnotesize{}%
\begin{tabular}{lccccccc}
\hline 
 &  & \multicolumn{3}{c}{$\tau=0.05$} & \multicolumn{3}{c}{$\tau=0.10$}\tabularnewline
\cline{3-8}
Design & $N=T$ & $\widehat{r}$ & $f_{3t}$ & RMSE & $\widehat{r}$ & $f_{3t}$ & RMSE\tabularnewline
\hline 
Independent & 100 & 2.89 & 0.926 & 0.693 & 2.93 & 0.924 & 0.571\tabularnewline
 & 200 & 3.00 & 0.976 & 0.479 & 3.00 & 0.973 & 0.397\tabularnewline
\hline 
(0.5, 0.5) & 100 & 2.96 & 0.836 & 0.888 & 2.96 & 0.800 & 0.760\tabularnewline
 & 200 & 3.00 & 0.924 & 0.625 & 3.00 & 0.909 & 0.538\tabularnewline
\hline 
\end{tabular}}{\footnotesize\par}
\par\end{centering}
{\footnotesize\emph{Notes:}}{\footnotesize{} 1,000 replications, $r=3$,
unit-variance errors; the dependent design has $(\phi,\gamma)=(0.5,0.5)$
and $J=3$. Column $f_{3t}$ reports the average adjusted $R^{2}$
for the scale factor; RMSE as in Table 2. Because the design is symmetric,
the results for $\tau=0.95$ and $0.90$ have the same distribution
as those reported.}{\footnotesize\par}
\end{table}

\subsection{Inference}

The inference experiment uses the design of CDG (Section 5.3), $X_{it}=\lambda_{i}f_{t}+f_{t}\varepsilon_{it}$,
in which $f_{t}\sim U(1,2)$, normalized so that $F'F/T=1$, and $\lambda_{i}\sim N(0,1)$
are held fixed across replications. The errors $\varepsilon_{it}$
follow Example 1 with unit variance and $J=3$. The quantile factor
model has one factor, $f_{t}$, with loading $\lambda_{i}+q_{\tau}$.
I compute the smoothed estimator with CDG's eighth-order kernel and
$h=T^{-0.15}$, which satisfies Assumption S, starting from QFA. The
Hessians are estimated with the Epanechnikov kernel and the bandwidth
of Corollary 2(b). For $\lambda_{0i}$, the estimator of Proposition
5 uses the Parzen kernel with the lag truncation $\max\{\widehat{B}_{T},T^{1/5}\}$
of Corollary 2. For $f_{0t}$, it uses the banded estimator with $J^{*}=2J=6$.
For the thresholded estimator of Proposition 6 I set $\omega_{NT}=\tau(1-\tau)(2\log N/T)^{1/2}$,
which is the universal threshold for the standard deviation $\tau(1-\tau)T^{-1/2}$
of $\widehat{\sigma}_{ij}$ under independence.

Theorem 2 indicates which intervals should fail. The asymptotic variance
of $\widehat{\lambda}_{i}$ involves the long-run variance of $\psi_{it}$
over $t$, and that of $\widehat{f}_{t}$ involves the long-run variance
over $i$. The variance estimators of CDG, which are designed for
independent errors, omit the autocovariances in the first and the
cross-covariances in the second. Serial dependence should therefore
cause undercoverage of the independence-based intervals for $\lambda_{0i}$
but not for $f_{0t}$, and cross-sectional dependence the reverse,
to an extent that depends on the loadings of dependent units, as I
discuss below. In both cases the shortfall should grow with the strength
of dependence. The estimators of Propositions 5 and 6 are consistent
for the long-run variances, so their intervals should have coverage
close to the nominal level, apart from the finite-sample bias of HAC
and banded estimators.

Table 6 reports the coverage of nominal 95 percent confidence intervals.
Under independence all estimators have coverage close to 0.95, except
the banded estimator for $f_{0t}$, which is estimated from a single
cross-section and has coverage of 0.90 at $N=100$ and 0.92 to 0.93
at $N=200$. Serial dependence leaves the intervals for $f_{0t}$
unaffected but invalidates CDG's intervals for $\lambda_{0i}$: their
coverage falls to about 0.82 when $\phi=0.5$ and to 0.59 when $\phi=0.8$,
because their standard errors are too small by factors of about 1.3
and 2. The HAC estimator of Proposition 5 raises coverage to between
0.91 and 0.93 when $\phi=0.5$ and to between 0.84 and 0.89 when $\phi=0.8$,
with the remaining shortfall due to the familiar downward bias of
HAC estimators in finite samples. Cross-sectional dependence has the
symmetric effect: with $\gamma=0.5$ and $\tau=0.25$, the coverage
of CDG's intervals for $f_{0t}$ falls to about 0.81, while the estimators
of Propositions 5 and 6 attain 0.90 to 0.92. These patterns match
the predictions of Theorem 2.

At the median, cross-sectional dependence leaves CDG's intervals for
$f_{0t}$ essentially valid. The cross-sectional long-run variance
$\Omega_{t}^{f}=\lim_{N}N^{-1}\sum_{i,j}\mathbb{E}[\psi_{it}\psi_{jt}]\lambda_{0i}\lambda_{0j}'$
differs from $\tau(1-\tau)\Sigma_{\Lambda}$ only through products
of the loadings of neighboring units. At $\tau=0.5$ the loadings
$\lambda_{i}$ are independent with mean zero, so these products average
out. At $\tau=0.25$ the loadings $\lambda_{i}+q_{\tau}$ share the
nonzero mean $q_{\tau}$, and the products do not average out. Cross-sectional
dependence in the errors therefore matters for inference on the factors
when the loadings of dependent units are correlated, which is the
empirically relevant case for, say, sectoral or regional panels. Finally,
the thresholded estimator of Proposition 6 loses accuracy when $\phi=0.8$
at the median, with coverage of 0.90 to 0.92. Serial dependence inflates
the sampling variability of $\widehat{\sigma}_{ij}$, so that a threshold
calibrated to independent data retains spurious entries. A threshold
based on the long-run standard deviation of $\psi_{it}\psi_{jt}$
would remove this effect.

\begin{table}[htbp]
\caption{\label{tab:mc-inf}Coverage of nominal 95\% confidence intervals}

\begin{centering}
{\footnotesize{}%
\begin{tabular}{lcccccccc}
\hline 
 &  &  & \multicolumn{2}{c}{$\lambda_{0i}$} & \multicolumn{3}{c}{$f_{0t}$} & \tabularnewline
\cline{4-8}
$N=T$ & $\tau$ & $(\phi,\gamma)$ & CDG & Prop. 5 & CDG & Prop. 5 & Prop. 6 & $\widehat{B}_{T}$\tabularnewline
\hline 
100 & 0.25 & (0, 0) & 0.946 & 0.941 & 0.944 & 0.896 & 0.944 & 4.1\tabularnewline
 &  & (0.5, 0) & 0.825 & 0.907 & 0.944 & 0.894 & 0.943 & 9.1\tabularnewline
 &  & (0.8, 0) & 0.591 & 0.842 & 0.941 & 0.888 & 0.942 & 18.1\tabularnewline
 &  & (0, 0.5) & 0.939 & 0.936 & 0.803 & 0.898 & 0.908 & 4.1\tabularnewline
 &  & (0.5, 0.5) & 0.815 & 0.903 & 0.801 & 0.896 & 0.905 & 9.0\tabularnewline
\hline 
100 & 0.50 & (0, 0) & 0.954 & 0.949 & 0.954 & 0.904 & 0.952 & 4.0\tabularnewline
 &  & (0.5, 0) & 0.826 & 0.919 & 0.951 & 0.900 & 0.945 & 9.1\tabularnewline
 &  & (0.8, 0) & 0.589 & 0.862 & 0.945 & 0.894 & 0.904 & 17.4\tabularnewline
 &  & (0, 0.5) & 0.944 & 0.938 & 0.939 & 0.903 & 0.938 & 4.0\tabularnewline
 &  & (0.5, 0.5) & 0.809 & 0.906 & 0.930 & 0.900 & 0.928 & 9.0\tabularnewline
\hline 
200 & 0.25 & (0, 0) & 0.948 & 0.946 & 0.944 & 0.915 & 0.944 & 3.8\tabularnewline
 &  & (0.5, 0) & 0.824 & 0.922 & 0.944 & 0.915 & 0.944 & 9.4\tabularnewline
 &  & (0.8, 0) & 0.590 & 0.880 & 0.943 & 0.914 & 0.943 & 18.2\tabularnewline
 &  & (0, 0.5) & 0.945 & 0.944 & 0.815 & 0.916 & 0.923 & 3.8\tabularnewline
 &  & (0.5, 0.5) & 0.818 & 0.920 & 0.814 & 0.916 & 0.922 & 9.4\tabularnewline
\hline 
200 & 0.50 & (0, 0) & 0.954 & 0.953 & 0.954 & 0.925 & 0.953 & 3.8\tabularnewline
 &  & (0.5, 0) & 0.823 & 0.931 & 0.953 & 0.923 & 0.949 & 9.7\tabularnewline
 &  & (0.8, 0) & 0.588 & 0.892 & 0.949 & 0.920 & 0.916 & 18.2\tabularnewline
 &  & (0, 0.5) & 0.951 & 0.948 & 0.937 & 0.933 & 0.948 & 3.8\tabularnewline
 &  & (0.5, 0.5) & 0.815 & 0.926 & 0.934 & 0.933 & 0.945 & 9.7\tabularnewline
\hline 
\end{tabular}}{\footnotesize\par}
\par\end{centering}
{\footnotesize\emph{Notes:}}{\footnotesize{} 1,000 replications of the
design of CDG (Section 5.3) with errors as in Example 1 ($J=3$, unit
variance). Entries are empirical coverage rates averaged over $i$
(for $\lambda_{0i}$) or $t$ (for $f_{0t}$). CDG: the estimators
of their Remark 4.4. Prop. 5: HAC estimator with the Parzen kernel
and the plug-in lag truncation $\max\{\widehat{B}_{T},T^{1/5}\}$
for $\lambda_{0i}$, and the banded estimator with $J^{*}=6$ for
$f_{0t}$. Prop. 6: thresholded estimator, which does not use the
ordering of the units. $\widehat{B}_{T}$ is the average lag truncation.}{\footnotesize\par}
\end{table}

I also check Proposition 7 by adding a common volatility path, $\sigma_{t}=2$
on 30 percent of the dates and 1 otherwise, that leaves the quantile
factor model unchanged (Table 7). The theory predicts that the intervals
of Propositions 5 and 6 keep their nominal coverage on calm and volatile
dates alike, that the standard errors of the factors are about twice
as large on volatile dates, because the asymptotic variance of $\widetilde{f}_{t}$
is proportional to $\sigma_{t}^{2}$, and that intervals replacing
$\widetilde{\Psi}_{t}$ by its average over time are too wide on calm
dates and too narrow on volatile ones. Table 7 supports all three
predictions: the thresholded intervals for $f_{0t}$ cover 0.91 to
0.97 on both kinds of dates, the HAC intervals for $\lambda_{0i}$
cover 0.93 to 0.96, the standard errors of the factors are 1.83 to
1.90 times larger on volatile dates, and the intervals based on the
average Hessian cover 0.96 to 0.98 on calm dates but only 0.71 to
0.77 on volatile ones.

\begin{table}[htbp]
\caption{\label{tab:mc-vol}Coverage under common volatility}

\begin{centering}
{\footnotesize{}%
\begin{tabular}{lccccccccc}
\hline 
 &  & $\lambda_{0i}$ & \multicolumn{2}{c}{$f_{0t}$: CDG} & \multicolumn{2}{c}{$f_{0t}$: Prop. 6} & \multicolumn{2}{c}{$f_{0t}$: pooled} & \tabularnewline
\cline{3-9}
$\tau$ & $(\phi,\gamma)$ & Prop. 5 & calm & volatile & calm & volatile & calm & volatile & SE ratio\tabularnewline
\hline 
0.25 & (0, 0) & 0.955 & 0.956 & 0.932 & 0.956 & 0.931 & 0.973 & 0.748 & 1.90\tabularnewline
 & (0.5, 0.5) & 0.928 & 0.830 & 0.794 & 0.934 & 0.907 & 0.957 & 0.710 & 1.90\tabularnewline
0.50 & (0, 0) & 0.961 & 0.966 & 0.939 & 0.965 & 0.937 & 0.977 & 0.771 & 1.83\tabularnewline
 & (0.5, 0.5) & 0.934 & 0.944 & 0.914 & 0.956 & 0.929 & 0.968 & 0.749 & 1.84\tabularnewline
\hline 
\end{tabular}}{\footnotesize\par}
\par\end{centering}
{\footnotesize\emph{Notes:}}{\footnotesize{} 1,000 replications with
$N=T=200$ of the design of Table 6 with $X_{it}=(\lambda_{i}+q_{\tau})f_{t}+f_{t}\sigma_{t}(\varepsilon_{it}-q_{\tau})$,
where $q_{\tau}$ is the $\tau$-quantile of $\varepsilon_{it}$,
$\sigma_{t}=2$ on 30 percent of the dates (volatile, in blocks of
six) and $\sigma_{t}=1$ otherwise (calm). Coverage of nominal 95
percent intervals. Pooled: Prop. 6 with $\widetilde{\Psi}_{t}$ replaced
by its average over $t$. SE ratio: average Prop. 6 standard error,
volatile relative to calm dates.}{\footnotesize\par}
\end{table}

In summary, the experiments support the theory of Section 3. Dependence
lowers the precision of QFA gradually, through the constant rather
than the rate. The threshold for the number of factors and the consistency
of the common component fail near the boundaries implied by Corollary
1 and Theorem 1. The rate $L_{NT}$ holds for all the error processes
considered, including some outside the Markov-driven class. Finally,
the variance estimators of Propositions 5 and 6 correct the substantial
undercoverage of intervals that assume independence, and they remain
accurate under common volatility.

\section{Empirical Illustration}

This section applies the results to the macroeconomic panels used
in the factor literature. I ask two questions that the theory makes
empirically relevant: whether the idiosyncratic errors of a standard
panel are consistent with the exact model, and whether allowing for
dependence changes inference.

I use the quarterly FRED-QD database (McCracken and Ng, 2020) over
1960Q1--2019Q2, the sample of CDG ($T=238$), and the monthly FRED-MD
database (McCracken and Ng, 2016) over 1960:01--2019:12 ($T=720$),
both from the September 2026 vintage. I end both samples before 2020
to exclude the pandemic observations. Each series is transformed to
stationarity with the codes supplied with the data, observations more
than ten interquartile ranges from the median of their series are
replaced by that median, and series with missing observations in the
sample are dropped. This leaves $N=209$ quarterly series and $N=121$
monthly series. All series are standardized. I estimate QFA at $\tau\in\{0.1,0.25,0.5,0.75,0.9\}$
and select $r(\tau)$ by the rank-minimization estimator with $k=8$,
as in Section 4.

The theory gives two predictions. Under the exact model the scores
$\widehat{\psi}_{it}=\tau-\mathbb{I}\{\widehat{u}_{it}\le0\}$ behave
approximately as independent mean-zero variables. About 5 percent
of the series should then have a first-order autocorrelation outside
$\pm1.96/\sqrt{T}$ or reject the Ljung--Box test at the 5 percent
level, and about 5 percent of the pairs of series should have a contemporaneous
correlation outside $\pm1.96/\sqrt{T}$. Under the approximate model
these fractions are larger. In that case the variance estimators of
Propositions 5 and 6 should give standard errors that differ from
those that assume independence, and that are larger when the dependence
is positive. For the factors, the sign of the dependence is measured
by the share of positive covariances among the score covariances retained
by the threshold of Proposition 6.

Table 8 rejects the exact model in both panels. In FRED-QD, between
40 and 68 percent of the series have a significant first-order score
autocorrelation, and between 52 and 70 percent reject the Ljung--Box
test. In FRED-MD the corresponding ranges are 64 to 90 percent and
74 to 92 percent. Cross-sectional dependence is also present: 7 to
15 percent of the pairs are significantly correlated in FRED-QD and
21 to 29 percent in FRED-MD, against 5 percent under independence.
The idiosyncratic errors of both panels are therefore dependent over
time and across series, which is the setting of the approximate model
rather than the exact one. Table 8 also shows that fewer factors are
selected in the tails: $\widehat{r}(\tau)$ is between 4 and 6 at
the central quantiles but only 1 or 2 at $\tau=0.1$ and $\tau=0.9$.

\begin{table}
\caption{\label{tab:emp-dep}Dependence in the QFA scores of FRED-QD and FRED-MD}

\begin{centering}
{\footnotesize{}%
\begin{tabular}{llccccc}
\hline 
 &  & $\tau=0.1$ & $0.25$ & $0.5$ & $0.75$ & $0.9$\tabularnewline
\hline 
FRED-QD & $\widehat{r}(\tau)$ & 2 & 4 & 5 & 5 & 2\tabularnewline
 & Autocorrelation & 0.40 & 0.58 & 0.68 & 0.55 & 0.42\tabularnewline
 & Ljung--Box & 0.52 & 0.59 & 0.70 & 0.53 & 0.52\tabularnewline
 & Cross-correlation & 0.07 & 0.14 & 0.15 & 0.14 & 0.07\tabularnewline
\hline 
FRED-MD & $\widehat{r}(\tau)$ & 1 & 5 & 6 & 5 & 1\tabularnewline
 & Autocorrelation & 0.64 & 0.77 & 0.90 & 0.84 & 0.64\tabularnewline
 & Ljung--Box & 0.74 & 0.88 & 0.92 & 0.91 & 0.74\tabularnewline
 & Cross-correlation & 0.22 & 0.29 & 0.21 & 0.26 & 0.22\tabularnewline
\hline 
\end{tabular}}{\footnotesize\par}
\par\end{centering}
{\footnotesize\emph{Notes:}}{\footnotesize{} Scores $\widehat{\psi}_{it}=\tau-\mathbb{I}\{\widehat{u}_{it}\le0\}$
from QFA with $\widehat{r}(\tau)$ factors, selected by the rank-minimization
estimator with $k=8$. Autocorrelation: fraction of series whose first-order
score autocorrelation exceeds $1.96/\sqrt{T}$ in absolute value.
Ljung--Box: fraction of series for which the Ljung--Box test with
four lags rejects at the 5 percent level. Cross-correlation: fraction
of pairs $(i,j)$, $i<j$, whose contemporaneous score correlation
exceeds $1.96/\sqrt{T}$ in absolute value. Under independence the
last three fractions are about 0.05. FRED-QD: 1960Q1--2019Q2, $N=209$,
$T=238$. FRED-MD: 1960:01--2019:12, $N=121$, $T=720$.}{\footnotesize\par}
\end{table}

Table 9 compares the standard errors. I compute the smoothed estimator
from QFA with the settings of Section 4.5 and report, for each quantile,
the ratio of the dependence-robust standard error to the independence-based
standard error of CDG (their Remark 4.4) across all loadings and across
all factors. For the loadings, the correction is moderate at the central
quantiles, with 50th percentiles of the ratio across loadings between
1.03 and 1.12, but it varies widely across series: the 25th and 75th
percentiles range from about 0.9 to about 1.4. In the tails the correction
is larger and more uniform. At $\tau=0.1$ and $\tau=0.9$ the 50th
percentile of the ratio is 1.11 to 1.13 in FRED-QD and 1.48 to 1.62
in FRED-MD, and the ratio exceeds one, so that the robust standard
error is larger than the independence-based one, for 66 to 100 percent
of the loadings. For the factors, the 50th percentile of the ratio
in FRED-QD is 1.26 to 1.33 at the central quantiles and 1.44 to 1.46
in the tails, and in the tails of FRED-MD it is 1.57 to 1.64. Intervals
that assume independence therefore overstate the precision of the
estimates, and they do so most in the tails, which are the quantiles
of main interest in applications to risk.

\begin{table}
\caption{\label{tab:emp-se}Ratio of dependence-robust to independence-based
standard errors}

\begin{centering}
{\footnotesize{}%
\begin{tabular}{llccccc}
\hline 
 &  & $\tau=0.1$ & $0.25$ & $0.5$ & $0.75$ & $0.9$\tabularnewline
\hline 
FRED-QD & $\widehat{\lambda}_{i}$: 50th pct. & 1.11 & 1.04 & 1.03 & 1.04 & 1.13\tabularnewline
 & $\widehat{\lambda}_{i}$: 25th--75th pct. & 0.96--1.34 & 0.91--1.21 & 0.87--1.23 & 0.90--1.22 & 0.94--1.41\tabularnewline
 & $\widehat{\lambda}_{i}$: share ratio $>1$ & 0.70 & 0.57 & 0.54 & 0.57 & 0.66\tabularnewline
 & $\widehat{f}_{t}$: 50th pct. & 1.46 & 1.33 & 1.30 & 1.26 & 1.44\tabularnewline
 & $\widehat{f}_{t}$: 25th--75th pct. & 1.21--1.61 & 1.29--1.40 & 1.25--1.32 & 1.22--1.28 & 1.23--1.57\tabularnewline
 & $\widehat{f}_{t}$: share $\widehat{\sigma}_{ij}>0$ & 1.00 & 0.73 & 0.58 & 0.70 & 1.00\tabularnewline
\hline 
FRED-MD & $\widehat{\lambda}_{i}$: 50th pct. & 1.62 & 1.12 & 1.05 & 1.09 & 1.48\tabularnewline
 & $\widehat{\lambda}_{i}$: 25th--75th pct. & 1.36--2.02 & 0.94--1.39 & 0.89--1.24 & 0.96--1.29 & 1.23--1.92\tabularnewline
 & $\widehat{\lambda}_{i}$: share ratio $>1$ & 1.00 & 0.66 & 0.57 & 0.67 & 0.94\tabularnewline
 & $\widehat{f}_{t}$: 50th pct. & 1.64 & 0.82 & 0.58 & 0.85 & 1.57\tabularnewline
 & $\widehat{f}_{t}$: 25th--75th pct. & --- & 0.73--0.93 & 0.53--0.68 & 0.77--0.96 & ---\tabularnewline
 & $\widehat{f}_{t}$: share $\widehat{\sigma}_{ij}>0$ & 0.85 & 0.44 & 0.47 & 0.45 & 0.84\tabularnewline
\hline 
\end{tabular}}{\footnotesize\par}
\par\end{centering}
{\footnotesize\emph{Notes:}}{\footnotesize{} Ratio of the dependence-robust
standard error to the independence-based standard error of CDG (Remark
4.4), both computed from the smoothed estimator with $\widehat{r}(\tau)$
factors. Loadings $\widehat{\lambda}_{i}$: HAC estimator of Proposition
5 with the Parzen kernel and lag truncation $\max\{\widehat{B}_{T},T^{1/5}\}$.
Factors $\widehat{f}_{t}$: thresholded estimator of Proposition 6
with $\omega_{NT}=\tau(1-\tau)(2\log N/T)^{1/2}$. The 50th and the
25th--75th percentiles are taken over all elements of $\widehat{\lambda}_{i}$,
$i\le N$, or of $\widehat{f}_{t}$, $t\le T$. Share ratio $>1$:
fraction of loadings whose robust standard error exceeds the independence-based
one. Share $\widehat{\sigma}_{ij}>0$: fraction of the off-diagonal
score covariances $\widehat{\sigma}_{ij}$ retained by the threshold
that are positive. With $\widehat{r}(\tau)=1$ the factor ratio is
the same for all $t$, so no percentile range is reported.}{\footnotesize\par}
\end{table}

The exception is the factors at the central quantiles of FRED-MD,
where the thresholded standard errors are smaller than the independence-based
ones, with 50th percentiles of the ratio between 0.58 and 0.85. The
last row of each panel of Table 9 shows why. In FRED-MD, only 44 to
47 percent of the retained covariances $\widehat{\sigma}_{ij}$ are
positive at $\tau\in\{0.25,0.5,0.75\}$, where the factor ratio is
below one, against 84 to 85 percent in the tails, where the ratio
is above one. In FRED-QD the share is 58 to 73 percent at the central
quantiles and 100 percent in the tails, and the ratio is above one
throughout. The factor ratio is therefore below one where the retained
score covariances are predominantly negative, as the prediction above
implies. Part of this negative dependence is likely induced by estimation.
The first-order condition of QFA for $\widehat{f}_{t}$ implies that
$N^{-1}\sum_{i}\widehat{\psi}_{it}\widehat{\lambda}_{i}\approx0$
for each $t$, so the unthresholded estimate $N^{-1}\sum_{i,j}\widehat{\sigma}_{ij}\widehat{\lambda}_{i}\widehat{\lambda}_{j}'$
of the cross-sectional long-run variance is close to zero: at $\tau=0.5$
in FRED-MD its trace is 0.01, against 0.10 for the independence-based
estimate $\tau(1-\tau)\widehat{\Lambda}'\widehat{\Lambda}/N$. Thresholding
removes most of this effect, but with five or six estimated factors
the constraint induces negative covariances among the scores, some
of which exceed the threshold. The data do not allow me to separate
this induced component from genuine negative dependence among the
idiosyncratic errors, and a correction of the estimator of Proposition
6 for the former is left for future research.

As a robustness check, I extend both samples to the end of the current
vintage, 2026Q2 for FRED-QD ($T=266$, $N=209$) and 2026:07 for FRED-MD
($T=799$, $N=120$). The extended samples include the pandemic observations,
which the outlier rule treats as above, and one-month gaps in several
FRED-MD series in October 2025, which I fill by linear interpolation.
The selected numbers of factors are the same as in Table 8 at every
quantile, and the fractions in Table 8 and the 50th percentiles of
the ratios in Table 9 change by at most 0.08.

In summary, the idiosyncratic errors of both FRED panels are serially
and cross-sectionally dependent, so the approximate model, not the
exact model, is the relevant framework for these data. Inference that
ignores this dependence understates uncertainty, most strongly in
the tails, where the robust standard errors of the factors are about
1.4 to 1.6 times those that assume independence in both panels.

\section{Conclusion}

This paper defines and analyzes the approximate quantile factor model.
The eigenvalue condition that defines approximate factor models in
the mean case does not control quantile estimation and is not nested
with the property that does (Proposition 2). I therefore formulate
weak dependence at a quantile as a concentration property of the idiosyncratic
array. Under this condition, QFA consistently estimates the factors,
the loadings and the number of factors at the rate $L_{NT}/\sqrt{C_{D}}$,
which is minimax optimal, and dependence changes the asymptotic variances,
for which I provide consistent HAC and thresholded estimators.

In the FRED-QD and FRED-MD panels, the idiosyncratic errors are dependent
over time and across series, and standard errors that ignore this
dependence understate uncertainty, most strongly in the tails. Two
extensions are natural: dynamic quantile factor models, in which the
factors follow a vector autoregression, and a correction of the thresholded
variance estimator for the negative dependence that estimation induces
among the scores.

\appendix

\section{Notation and Proofs}

\subsection{Notation}

The table collects the notation used in Section 3 and in this appendix;
the additional notation for the distribution theory of Section 3.4
is collected in the Supplemental Appendix. Unless stated otherwise,
probabilities and expectations are conditional on $F_{0}$, and $\Vert\cdot\Vert$
is the Euclidean norm for vectors and the Frobenius norm for matrices.

\noindent{}%
\begin{longtable}[l]{>{\raggedright}p{3.3cm}>{\raggedright}p{11.1cm}}
\hline 
\textbf{Symbol} & \textbf{Description}\tabularnewline
\hline 
\endfirsthead
\hline 
\textbf{Symbol} & \textbf{Description}\tabularnewline
\hline 
\endhead
\multicolumn{2}{l}{\textbf{Model and parameters}}\tabularnewline
$x_{it}$ & Observation for unit $i=1,\dots,N$ in period $t=1,\dots,T$.\tabularnewline
$\tau$ & Quantile level, fixed throughout.\tabularnewline
$r$; $l$, $k_{\max}$ & Number of factors at quantile $\tau$; candidate numbers of factors
in Corollary 1.\tabularnewline
$\lambda_{i}$, $f_{t}$; $\Lambda$, $F$ & Loadings and factors; $\Lambda=(\lambda_{1},\dots,\lambda_{N})'$
and $F=(f_{1},\dots,f_{T})'$.\tabularnewline
$\theta$, $\theta_{0}$ & Parameter $(\lambda_{1}',\dots,\lambda_{N}',f_{1}',\dots,f_{T}')'$
and its true value; $\Lambda_{0}$, $F_{0}$ are the true matrices.\tabularnewline
$\Theta^{r}$, $\Theta^{l}$ & Parameter space with $r$ (or $l$) factors: $\Vert\lambda_{i}\Vert,\Vert f_{t}\Vert\le B$,
$F'F/T=I$, $\Lambda'\Lambda/N$ diagonal and non-increasing (A1).\tabularnewline
$B$ & Bound on the norms of loadings and factors.\tabularnewline
$\sigma_{Nk}$, $\sigma_{k}$, $\Sigma_{\Lambda}$ & Diagonal elements of $\Lambda_{0}'\Lambda_{0}/N$, their limits, and
$\Sigma_{\Lambda}=\mathrm{diag}(\sigma_{1},\dots,\sigma_{r})$ (pervasiveness
(P)).\tabularnewline
$u_{it}(\tau)$, $\varepsilon_{it}$ & Idiosyncratic error at quantile $\tau$; $\varepsilon_{it}=u_{it}(\tau)$,
with $P(\varepsilon_{it}\le0\mid F_{0})=\tau$.\tabularnewline
$\rho_{\tau}$ & Check function, $\rho_{\tau}(u)=u(\tau-\mathbb{I}\{u\le0\})$.\tabularnewline
$\psi_{it}$ & Check-function score, $\psi_{it}=\tau-\mathbb{I}\{u_{it}\le0\}$.\tabularnewline
$F_{it}$, $f_{it}$ & Distribution function and density of $\varepsilon_{it}$ given $F_{0}$.\tabularnewline
$\underline{f}$, $\kappa$ & Lower bound of the density and half-width of the neighborhood of zero
on which it holds (A2).\tabularnewline
$L_{NT}$, $\widetilde{L}_{NT}$ & $L_{NT}=\min\{\sqrt{N},\sqrt{T}\}$ and the dependence-adjusted rate
$\widetilde{L}_{NT}=L_{NT}/\sqrt{C_{D}}$.\tabularnewline
$\widehat{S}$, $\widetilde{S}$ & Diagonal sign matrices aligning estimated and true factors, with $\mathrm{sgn}(0)=1$.\tabularnewline
\multicolumn{2}{l}{\textbf{Criterion and metrics}}\tabularnewline
$\delta_{it}(\theta)$ & Deviation of the common component, $\lambda_{i}'f_{t}-\lambda_{0i}'f_{0t}$.\tabularnewline
$X(\theta)$ & Scaled common component $\Lambda F'/\sqrt{NT}$.\tabularnewline
$d(\theta,\theta')$ & Frobenius distance $\Vert X(\theta)-X(\theta')\Vert$, the root mean
squared difference of common components.\tabularnewline
$d_{\infty}(\theta,\theta')$ & Maximum distance $\max_{i,t}|\delta_{it}(\theta)-\delta_{it}(\theta')|$
(proof of Proposition 4(b)).\tabularnewline
$M(\theta)$, $\mathbb{M}(\theta)$ & QFA criterion $(NT)^{-1}\sum_{i,t}\rho_{\tau}(\varepsilon_{it}-\delta_{it}(\theta))$
and its conditional expectation given $F_{0}$.\tabularnewline
$G(\theta)$ & Centered criterion increment, $NT[M(\theta)-M(\theta_{0})]-NT[\mathbb{M}(\theta)-\mathbb{M}(\theta_{0})]$.\tabularnewline
$\Theta(\delta)$ & Ball $\{\theta\in\Theta^{r}:d(\theta,\theta_{0})\le\delta\}$.\tabularnewline
$c_{0}$ & Curvature constant of Lemma B.1, $c_{0}=\underline{f}\min\{1/2,\kappa/(4B^{2})\}$.\tabularnewline
$\eta_{NT}$ & Rate sequence $\sqrt{C_{D}r(N+T)/(NT)}$ in the proof of Theorem 1
(Supplemental Appendix B).\tabularnewline
$\widehat{d}$ & Estimation error $d(\widehat{\theta},\theta_{0})$.\tabularnewline
$\widehat{d}_{j}$ & $j$th diagonal element of $\widehat{\Lambda}'\widehat{\Lambda}/N$.\tabularnewline
$P_{NT}$; $\widehat{r}_{\mathrm{rank}}$, $\widehat{r}_{\mathrm{IC}}$ & Threshold or penalty; rank-minimization and information-criterion
estimators of $r$.\tabularnewline
\multicolumn{2}{l}{\textbf{Dependence}}\tabularnewline
$C_{D}$ & Concentration constant in A3; weak dependence means $C_{D}=O(1)$.\tabularnewline
$g_{it}$, $c_{it}$ & Bounded functions in A3 and bounds on their oscillation.\tabularnewline
$H$ & Number of classes of mutually independent units, D(a).\tabularnewline
$Z_{it}$, $\varphi_{it}$ & Markov chain driving unit $i$ and the map with $\varepsilon_{it}=\varphi_{it}(Z_{it})$,
D(b).\tabularnewline
$\beta_{i}(k)$, $\alpha_{i}(k)$ & Absolute regularity and strong mixing coefficients of unit $i$'s
errors at lag $k$, D(c)--(d).\tabularnewline
$C_{\beta}$, $c_{\beta}$, $c_{\alpha}$ & Constants in the geometric bounds of D(c) and D(d).\tabularnewline
$q_{NT}$ & Block length $\lceil2\log(NT)/c_{\beta}\rceil$ in Proposition 4(a).\tabularnewline
$\beta(k)$; $C_{A}$, $\varsigma$, $a$ & $\beta$-mixing coefficient of the error vector $u_{t}$ and the constants
in $\beta(k)\le C_{A}N^{\varsigma}a^{k}$, A4(a).\tabularnewline
$J^{*}$ & Range of cross-sectional dependence, A4(b).\tabularnewline
$k_{NT}$ & Lag cutoff $\lceil(\varsigma+3)\log(NT)/\log(1/a)\rceil$, with $\beta(k_{NT})\le C_{A}(NT)^{-3}$.\tabularnewline
\multicolumn{2}{l}{\textbf{Constants}}\tabularnewline
$K$ & Absolute constant in Dudley's bound (Lemma B.3).\tabularnewline
$K_{1}$ & Absolute constant converting a sub-Gaussian tail into a moment generating
function bound.\tabularnewline
$M_{0}$, $M_{\phi}$ & Constants in the inequality of Dedecker and Gouëzel (2015, Theorem
2), for a given transition kernel.\tabularnewline
$C$, $c$ & Generic positive constants that do not depend on $i$, $t$, $N$
or $T$; their values may change between uses.\tabularnewline
\hline 
\end{longtable}

\subsection{Proof of Theorem 3}

\emph{Model.} For every $m$, each $u_{it}$ has the law of $s(z-\Phi^{-1}(\tau))$
with $z\sim N(0,1)$, so $P(u_{it}\le0)=\tau$ and the density of
$u_{it}$ is continuous and bounded away from zero near zero, which
gives A2. A1 holds for the hypotheses constructed below. For A3, partition
the index set $\{(i,t)\}$ into the blocks on which the errors coincide.
Each block has at most $m$ elements, and the errors are independent
across blocks. Given functions $g_{it}$ with oscillations $c_{it}$,
the sum of the $g_{it}(\varepsilon_{it})$ over a block $b$ is a
function of the common value of the errors in $b$ with oscillation
at most $c_{b}=\sum_{(i,t)\in b}c_{it}$, and $c_{b}^{2}\le m\sum_{(i,t)\in b}c_{it}^{2}$
by the Cauchy--Schwarz inequality. Hoeffding's inequality applied
to the independent block sums therefore gives A3 with $C_{D}=m/2$.
With $m=1$ the errors are independent and $C_{D}=1/2$.

Suppose $N\le T$, so that $L_{NT}=\sqrt{N}$ and the blocks consist
of $m$ consecutive units. The case $N>T$ is symmetric, with the
roles of units and periods exchanged, $C(\omega)=(\mathbf{1}_{N}+\varpi\omega)\mathbf{1}_{T}'$
for $\omega\in\{-1,1\}^{N}$, $f_{t}=1$, $\lambda_{i}=1+\varpi\omega_{i}$,
and $T$ in place of $N$ in the choice of $\varpi$ below.

\emph{Hypotheses.} By the Varshamov--Gilbert bound (Tsybakov, 2009,
Lemma 2.9), for $T\ge8$ there is a set $\Omega\subset\{-1,1\}^{T}$
with $M=|\Omega|\ge2^{T/8}$ whose elements differ pairwise in at
least $T/8$ coordinates. For $\omega\in\Omega\cup\{0\}$, let $C(\omega)=\mathbf{1}_{N}(\mathbf{1}_{T}+\varpi\omega)'$
with $\varpi\le1/2$. Each $C(\omega)$ has rank one and entries in
$[1/2,3/2]$. Write $C(\omega)=\Lambda F'$ with $\lambda_{i}=c(\omega)=\{T^{-1}\sum_{t}(1+\varpi\omega_{t})^{2}\}^{1/2}\in[1/2,3/2]$
and $f_{t}=(1+\varpi\omega_{t})/c(\omega)$. Then $F'F/T=1$, $\Lambda'\Lambda/N=c(\omega)^{2}\ge1/4$,
and $|\lambda_{i}|,|f_{t}|\le3\le B$, so $(\Lambda,F)\in\Theta^{1}$.

\emph{Separation.} For distinct $\omega,\omega'\in\Omega$, $(NT)^{-1}\Vert C(\omega)-C(\omega')\Vert^{2}=(\varpi^{2}/T)\sum_{t}(\omega_{t}-\omega_{t}')^{2}\ge\varpi^{2}/2$.
Also $(NT)^{-1}\Vert C(\omega)-C(0)\Vert^{2}=\varpi^{2}$. Hence all
hypotheses are separated by at least $2\delta$ with $\delta=\varpi/(2\sqrt{2})$,
the separation radius, which is distinct from the noise scale $s$
of the theorem.

\emph{Kullback--Leibler divergence.} Under $C(\omega)$ the rows
of the mean matrix are identical, so the observations within each
block of units are identical copies of a single row. The data are
therefore equivalent to $n_{m}=\lceil N/m\rceil\le2N/m$ independent
rows, each consisting of $T$ independent normal observations with
means $C(\omega)_{1t}$ (up to a common shift) and variance $s^{2}$.
The common support of these laws does not depend on $\omega$, so
the Kullback--Leibler divergence equals that of the laws of the block
rows, and $\mathrm{KL}(P_{\omega},P_{0})=n_{m}T\varpi^{2}/(2s^{2})\le NT\varpi^{2}/(ms^{2})$,
where $n_{m}\le2N/m$ uses $m\le N$, which holds because $c_{s}\le1$.
With $\varpi^{2}=s^{2}m\log2/(128N)$, which is at most $1/4$ because
$m\le c_{s}N$, this is at most $T\log2/128\le\frac{1}{16}\log M$.

\emph{Conclusion.} Theorem 2.5 of Tsybakov (2009), with $\alpha_{\mathrm{F}}=1/16<1/8$,
gives $\inf_{\widehat{C}}\max_{\omega\in\Omega\cup\{0\}}P_{\omega}\big((NT)^{-1/2}\Vert\widehat{C}-C(\omega)\Vert\ge\delta\big)\ge c'>0$
for $T\ge8$. Since $\delta=\varpi/(2\sqrt{2})=s\sqrt{m\log2}/(32\sqrt{N})$
and $m=2C_{D}$, $\delta=c\sqrt{C_{D}}/L_{NT}$ with $c=s\sqrt{2\log2}/32$,
and the claim follows. $\blacksquare$
\end{document}